\documentclass[aps,prl,reprint,superscriptaddress,nofootinbib,longbibliography]{revtex4-2}
\usepackage{etoolbox,graphicx,amsmath,amssymb,bm,mathrsfs}
\usepackage[Symbol]{upgreek}
\usepackage[hidelinks,hypertexnames=false]{hyperref}

\begin{document}
\title{Orbital-Selective Diagonal-Gap Test of Pairing in La$_3$Ni$_2$O$_7$}
\author{Yu-Bo Liu}\thanks{These authors contributed equally.}
\affiliation{College of Sciences, Northeastern University, Shenyang 110819, China}
\author{Zhi-Yan Shao}\thanks{These authors contributed equally.}
\affiliation{School of Physics, Beijing Institute of Technology, Beijing 100081, China}
\author{Zhiming Pan}
\affiliation{Department of Physics, Xiamen University, Xiamen 361005, Fujian, China}
\author{Chen Lu}
\affiliation{School of Physics, Hangzhou Normal University, Hangzhou 311121, China}
\author{Fan Yang}\email{yangfan_blg@bit.edu.cn}
\affiliation{School of Physics, Beijing Institute of Technology, Beijing 100081, China}
\begin{abstract}
Recent angle-resolved photoemission spectroscopy (ARPES) and scanning tunneling microscopy (STM) measurements on bilayer nickelate La$_3$Ni$_2$O$_7$ reveal a nearly isotropic, nodeless superconducting gap. We show that these single-particle spectra provide a symmetry-enforced test of the pairing nature. Mirror symmetry forces the hybridization and interorbital pairing (for the $s$-wave) between $d_{z^2}$ and $d_{x^2-y^2}$ orbitals to vanish along the Brillouin-zone (BZ) diagonal. Consequently, the gaps on the diagonal portions of the $\alpha/\beta$ and $\gamma$ Fermi pockets separately probe the intrinsic pairing strengths of the $d_{x^2-y^2}$ and $d_{z^2}$ sectors. A pairing state without an intrinsic $d_{x^2-y^2}$-orbital pair field produces nodes on the BZ diagonal of the $\alpha/\beta$ pockets and is thus challenged by ARPES and STM, whereas a pairing state including an appreciable intrinsic $d_{x^2-y^2}$-orbital pairing component can have a nearly isotropic, nodeless gap on all pockets, which is well consistent with the ARPES and STM data. The necessary $d_{x^2-y^2}$-orbital pairing component is compatible with Hund's-rule-driven pairing mechanisms.
\end{abstract}
\maketitle
\section*{Introduction}
The discovery of high-temperature superconductivity (SC) in bilayer nickelates La$_3$Ni$_2$O$_7$~\cite{Wang2023LNO,YuanHQ2023LNO,Wang2023LNOb} has stimulated intense studies of Ruddlesden-Popper phase nickelates~\cite{wang2023la2prnio7,liu2025superconductivity}, including their structure~\cite{wang2023structure,geisler2023structural,bhatt2025resolving,PhysRevB.112.L140504}, physical properties~\cite{Li2024ele,yang2024orbital,chen2024electronic,YaoDX2023,geisler2023structural,Yubo_Liu2024,li2025photoemission,wang2025electronic,sun2025observation,yue2025correlated,shao2025band,Daoxin_Yao2025,ushio2025theoretical,cao2025strain,shao2025pairing,shi2025effect,PhysRevB.112.L140504,bheemavarapu2025strain,wang2025origin,yang2025magnetism}, and pairing mechanism~\cite{shen2025anomalous,fan2025superconducting,wang2026atomically,liang2026observation,miao2026thermodynamic,li2025enhanced,10.1093/nsr/nwag151,YangF2023,zhang2023structural,zhang2023trends,gao2025robust,yue2025correlated,shao2025band,ryee2025optimal,shao2025pairing,heier2023competing,PhysRevB.111.104505,ushio2025theoretical,hua2026possible,WangQH2023,HuJP2023,cao2025strain,jiang2023high,fan2023sc,liao2023electron,jiang2023pressure,10.1093/nsr/nwaf353,Yi_Feng2023,ZhangGM2023DMRG,kaneko2023pair,WuWei2023charge,chen2026evolution,lu2023bilayertJ,oh2023type2,qu2023bilayer,zhang2023strong,yang2023strong,wu2024deconfined,pan2023rno,Lu2024interplay,PhysRevB.113.174521,kaneko2025tj,Ji2025StrongCouplingLimit,chen2026unified,liu2026triplon,shao_electric_field,fan2025minimal,qiu2025pairing}. The bilayer structure with strong interlayer hybridization together with the coexisting low-energy Ni-$3d_{z^2}$ and Ni-$3d_{x^2-y^2}$ orbitals differentiate this family from the cuprates. The different fillings of the two $E_g$-orbitals and the orbital-selective strong band renormalization revealed by the angle-resolved-photoemission-spectroscopy (ARPES)~\cite{yang2024orbital,Li2024ele} imply the different roles played by the two orbitals in the pairing mechanism. These features suggest that the orbital character of the pairing nature might be crucial for the pairing mechanism. Specifically, which orbital pairing components are essential for understanding the observed superconducting gap in La$_3$Ni$_2$O$_7$?

The proposed pairing states in bilayer nickelates include interlayer $s$-wave pairing~\cite{lu2023bilayertJ,oh2023type2,qu2023bilayer,zhang2023strong,yang2023strong,wu2024deconfined,Lu2024interplay,PhysRevB.113.174521,kaneko2025tj,Ji2025StrongCouplingLimit,shao2025pairing,liu2026triplon,Yi_Feng2023,ZhangGM2023DMRG,kaneko2023pair,WuWei2023charge,chen2026evolution} and intralayer $d$-wave one~\cite{jiang2023high,fan2023sc,liao2023electron,jiang2023pressure,10.1093/nsr/nwaf353}. Here we focus on $s$-wave pairing, which is compatible with the nodeless gap indicated by recent ARPES~\cite{shen2025anomalous,miao2026thermodynamic} and scanning tunneling microscopy (STM)~\cite{fan2025superconducting,wang2026atomically,liang2026observation} measurements. One strong-coupling viewpoint originates from the insight that the strong antiferromagnetic (AFM) interlayer superexchange between the $d_{z^2}$ electrons drives their interlayer $s$-wave pairing through orbital hybridization~\cite{Yi_Feng2023,ZhangGM2023DMRG,kaneko2023pair,WuWei2023charge,chen2026evolution}. By contrast, another strong-coupling viewpoint emphasizes the importance of the Hund's-rule, which transfers the interlayer AFM superexchange from the nearly-localized $d_{z^2}$ orbital to the more itinerant $d_{x^2-y^2}$ orbital~\cite{lu2023bilayertJ,oh2023type2,qu2023bilayer,zhang2023strong,yang2023strong,wu2024deconfined,Lu2024interplay,PhysRevB.113.174521,kaneko2025tj,Ji2025StrongCouplingLimit,shao2025pairing}, leading to their interlayer $s$-wave pairing. Alternatively, weak-coupling approaches such as the random-phase-approximation (RPA)~\cite{YangF2023,zhang2023structural,zhang2023trends,gao2025robust,yue2025correlated,shao2025band,ryee2025optimal,shao2025pairing}, the fluctuation-exchange (FLEX)~\cite{heier2023competing,PhysRevB.111.104505,ushio2025theoretical,hua2026possible} and the functional-renormalization-group (FRG)~\cite{WangQH2023,HuJP2023,cao2025strain} usually yield $s^\pm$-wave SC mediated by spin fluctuations. In these weak-coupling results, both orbital pairing components coexist, and the interlayer $d_{z^2}$-orbital pairing component is usually larger~\cite{YangF2023,gao2025robust,yue2025correlated,ushio2025theoretical,WangQH2023,HuJP2023,cao2025strain}.  Since all these orbital-selective scenarios lead to the same global $s$-wave gap symmetry, identifying the orbital character requires an orbital-resolved test of the pairing state.

The recent spectroscopic progress in La$_3$Ni$_2$O$_7$ thin films provides an opportunity to clarify this orbital character. ARPES measurements have resolved the pairing gap on the observed Fermi surfaces (FSs) of La$_3$Ni$_2$O$_7$, revealing nearly isotropic, nodeless gap~\cite{shen2025anomalous,miao2026thermodynamic}. In parallel, STM measurements show U-shaped tunneling spectra with a fully opened low-energy gap~\cite{fan2025superconducting,wang2026atomically,liang2026observation}. Together, these single-particle spectra support a fully gapped $s$-wave pairing state and provide the experimental basis for testing the orbital character of the pairing.

In this work, motivated by these spectroscopic observations, we show that the pairing gap along the Brillouin-zone (BZ) diagonal provides an orbital-resolved test of the pairing state. Because the $d_{z^2}$ and $d_{x^2-y^2}$ orbitals have opposite mirror parities, their hybridization and interorbital pairing (for the $s$-wave) vanish on the diagonal. Consequently, the diagonal gaps on the $\gamma$ and $\alpha/\beta$ pockets separately probe the intrinsic intraorbital pairing strength in the $d_{z^2}$ and $d_{x^2-y^2}$ sectors. Applying this test to the ARPES and STM spectra, we find that intrinsic pairing in the $d_{z^2}$ orbital alone leaves nodes on the $\alpha/\beta$ pockets along the BZ diagonal, and thus identify the $d_{x^2-y^2}$-orbital pairing component as essential for the observed nodeless diagonal gap. Including an appreciable $d_{x^2-y^2}$-orbital pair field, we well reproduce the measured spectroscopic data, further supporting the indispensable role of the $d_{x^2-y^2}$-orbital pairing component in La$_3$Ni$_2$O$_7$. Such a pairing state is compatible with Hund's-rule-driven pairing mechanisms.

\section*{Results}
\subsection*{Symmetry-enforced orbital-selective pairing}
To describe the low-energy band structure of La$_3$Ni$_2$O$_7$, we adopt the two-orbital tight-binding (TB) model~\cite{Daoxin_Yao2025}, whose Fermi-surface geometry closely resembles that observed by ARPES~\cite{shen2025anomalous,miao2026thermodynamic},
\begin{eqnarray}\label{H0}
H_{0} &=& H_0^x + H_0^z + H_0^{xz},
\end{eqnarray}
where $H_0^x$ ($H_0^z$) denotes the intra-$d_{x^2-y^2}$ ($d_{z^2}$) orbital component and $H_0^{xz}$ indicates the orbital hybridization, 
\begin{equation}\label{Hybridization_H}
H_{0}^{xz} = \sum_{\mathbf{k}\mu\nu\sigma} h_{\mu\nu}^{xz}(\mathbf{k})c_{\mathbf{k}\mu x\sigma}^\dagger c_{\mathbf{k}\nu z\sigma}+\text{H.c.}
\end{equation}
Here $\sigma$ labels spin, $\mu,\nu$ label layer, and $h_{\mu\nu}^{xz}$ is the interorbital hybridization matrix element.

\begin{figure}[tp]
    \centering
    \includegraphics[width=\columnwidth,keepaspectratio]{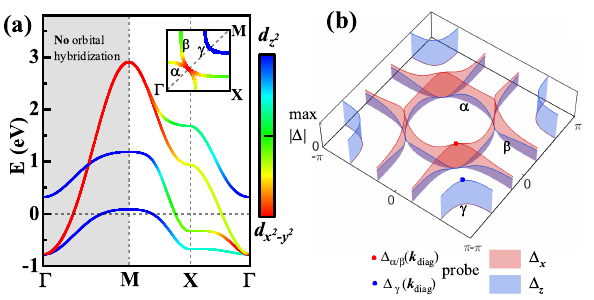}
    \caption{Orbital-selective diagonal-gap test. (a) Orbital-resolved band structure and Fermi surfaces (inset) of the two-orbital TB model of Ref.~\cite{Daoxin_Yao2025}. The color indicates the orbital weight. Along the Brillouin-zone diagonal, mirror symmetry forbids the orbital hybridization. (b) Gap distribution on the FS, obtained by projecting representative $d_{z^2}$- and $d_{x^2-y^2}$-orbital pair fields, $\Delta^z_{\perp}$ (blue) and $\Delta^x_{\perp}$ (red), onto the FSs. The diagonal gaps on the $\alpha/\beta$ pockets probe $\Delta^x_{\perp}$, while the diagonal gap on the $\gamma$ pocket probes $\Delta^z_{\perp}$.} \label{fig1}
\end{figure}

Note that SC in La$_3$Ni$_2$O$_7$ is associated with the $I4/mmm$ or $Fmmm$ phase, either in pressurized bulk or in compressively strained films, which possesses mirror-reflection symmetry about the in-plane diagonals~\cite{wang2023structure,bhatt2025resolving}. This mirror symmetry survives correlation-induced band renormalization~\cite{yang2024orbital}. Under this mirror reflection $\hat{M}$, the $d_{z^2}$ orbital is even, whereas the $d_{x^2-y^2}$ orbital is odd. Therefore, the $\hat{M}$-invariance of the system requires $h_{\mu\nu}^{xz}(\hat{M}\mathbf{k})=-h_{\mu\nu}^{xz}(\mathbf{k})$. For momenta on the Brillouin-zone (BZ) diagonal, we have $\mathbf{k}_{\rm diag}=\hat{M}\mathbf{k}_{\rm diag}$. Therefore,
\begin{eqnarray}\label{H0_diag}
h_{\mu\nu}^{xz}(\mathbf{k}_{\rm diag})=0, H_{0}(\mathbf{k}_{\rm diag}) =H_0^x(\mathbf{k}_{\rm diag}) + H_0^z(\mathbf{k}_{\rm diag}).
\end{eqnarray}
Here $H_0^{x/z}(\mathbf{k}_{\rm diag})$ denotes the $\mathbf{k}_{\rm diag}$-component of $H_0^{x/z}$. Eq.~(\ref{H0_diag}) shows that the normal-state eigenstates on the diagonal each carry a definite orbital component. This orbital decoupling is directly visible in Fig.~\ref{fig1}(a). Along the $\Gamma$-$M$ diagonal cut, two nearly degenerate $d_{x^2-y^2}$ bands and one $d_{z^2}$ band cross the Fermi level and form the diagonal portions of the $\alpha/\beta$- and $\gamma$- pockets shown in the inset, respectively.

The pairing state in La$_3$Ni$_2$O$_7$ can be generally described by the following Hamiltonian in the mean-field (MF) level,
\begin{equation}\label{eq:H_MF_general}
H_{\rm MF}=H_0+H^x_{\rm pair}+H^z_{\rm pair}+H^{xz}_{\rm pair}.
\end{equation}
Here $H^x_{\rm pair}$ ($H^z_{\rm pair}$) indicates the MF pair field Hamiltonian of the intrinsic $d_{x^2-y^2}$- ($d_{z^2}$-) orbital component. The $H^{xz}_{\rm pair}$ describes interorbital pairing and can be written as
\begin{equation}\label{eq:pair_mirror}
H^{xz}_{\rm pair}=\sum_{\mathbf{k}\mu\nu}\Delta^{xz}_{\mu\nu}(\mathbf{k})\left(c^\dagger_{\mathbf{k}\mu x\uparrow}c^\dagger_{-\mathbf{k}\nu z\downarrow}-c^\dagger_{\mathbf{k}\mu x\downarrow}c^\dagger_{-\mathbf{k}\nu z\uparrow}\right)+\text{H.c.}
\end{equation}
Under the mirror reflection $\hat{M}$, the $d_{z^2}$ orbital is even, whereas the $d_{x^2-y^2}$ orbital is odd. Therefore, for mirror-preserving $s$-wave pairing consistent with experiments, the $\hat{M}$-invariance of $H^{xz}_{\rm pair}$ requires $\Delta^{xz}_{\mu\nu}(\hat{M}\mathbf{k})=-\Delta^{xz}_{\mu\nu}(\mathbf{k})$. Since $\hat{M}\mathbf{k}_{\rm diag}=\mathbf{k}_{\rm diag}$ on the BZ diagonal, $\Delta^{xz}_{\mu\nu}(\mathbf{k}_{\rm diag})=0$. Therefore, for the $s$-wave pairing relevant to La$_3$Ni$_2$O$_7$, the interorbital pairing vanishes on the BZ diagonal.

This symmetry-enforced orbital decoupling on the diagonal, including the orbital dehybridization and the vanishing of interorbital pairing, allows the diagonal superconducting gap to probe the orbital character of the pairing state. On the BZ diagonal, since $H_0^{xz}(\mathbf{k}_{\rm diag})=0$ and $\Delta^{xz}_{\mu\nu}(\mathbf{k}_{\rm diag})=0$, Eq.~(\ref{eq:H_MF_general}) reduces to
\begin{eqnarray}\label{eq:H_MF_diag}
H_{\rm MF}(\mathbf{k}_{\rm diag}) &=& \left[H_0^x(\mathbf{k}_{\rm diag})+H^x_{\rm pair}(\mathbf{k}_{\rm diag})\right]\nonumber\\
&\oplus&\left[H_0^z(\mathbf{k}_{\rm diag})+H^z_{\rm pair}(\mathbf{k}_{\rm diag})\right].
\end{eqnarray}
Eq.~(\ref{eq:H_MF_diag}) suggests that on the BZ diagonal, the full MF Hamiltonian is block diagonalized into the $d_{x^2-y^2}$- sector and the $d_{z^2}$- sector. Since the $d_{x^2-y^2}$- ($d_{z^2}$-) orbital component is fully carried by the $\alpha/\beta$- ($\gamma$-) pocket on the diagonal, the pairing gap there fully reflects the effect of $H^x_{\rm pair}$ ($H^z_{\rm pair}$) and thus probes the intrinsic pairing strengths of the $d_{x^2-y^2}$ ($d_{z^2}$) sector. Since the orbital decoupling on the BZ diagonal is enforced by the mirror reflection symmetry, the resultant orbital-selective diagonal-gap criterion is insensitive to band details.

To thoroughly compare with ARPES/STM spectra, we project the pair field onto the FSs while retaining intraband pairing to obtain the gap distribution over the FSs (see Methods).  
For clarity, Fig.~\ref{fig1}(b) shows only the usually obtained interlayer $s$-wave intraorbital pairing components~\cite{PhysRevB.111.104505,ushio2025theoretical,qiu2025pairing,YangF2023,yue2025correlated,shao2025band,cao2025strain,WangQH2023,Lu2024interplay}, with amplitudes $\Delta^x_{\perp}$ and $\Delta^z_{\perp}$ for the $d_{x^2-y^2}$ and $d_{z^2}$ orbitals, respectively. Figure~\ref{fig1}(b) shows the projections of $\Delta^x_{\perp}$ and $\Delta^z_{\perp}$ onto the three pockets as the red and blue shaded surfaces, respectively, and their upper boundaries represent the corresponding gap magnitudes. At the diagonal Fermi momenta marked by the colored dots, the gaps on the $\alpha/\beta$ pockets arise entirely from the $d_{x^2-y^2}$-orbital pairing component, whereas the gap on the $\gamma$ pocket is determined solely by the $d_{z^2}$-orbital pairing component. Away from the diagonal, orbital hybridization allows both projected pairing components to contribute to each pocket.

The orbital-selective diagonal gaps therefore provide a direct criterion for determining the orbital character of the pairing. In the following, the calculated gaps and tunneling spectrum of representative pairing states are compared with ARPES/STM spectra to identify the essential orbital pairing components in La$_3$Ni$_2$O$_7$.

\begin{figure*}[tp]
    \centering
    \includegraphics[width=\textwidth,keepaspectratio]{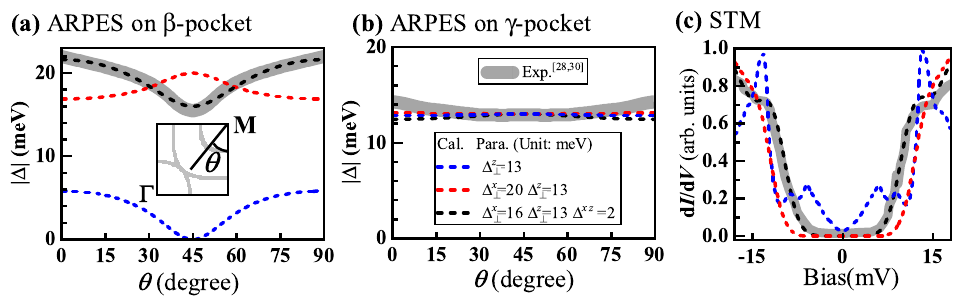}
    \caption{Comparison of the calculated superconducting gaps and tunneling spectra with ARPES and STM measurements. (a,b) Gap magnitudes on (a) the $\beta$ pocket and (b) the $\gamma$ pocket as functions of the angle $\theta$. The gray regions indicate the ARPES results~\cite{shen2025anomalous}. The blue dashed curves are calculated with only an interlayer $d_{z^2}$-orbital pair field, $\Delta^z_{\perp}=13\,\mathrm{meV}$, whereas the red dashed curves additionally include a $d_{x^2-y^2}$-orbital pair field, $\Delta^x_{\perp}=20\,\mathrm{meV}$. The black dashed curves show the minimal fits including nearest-neighbor intralayer interorbital pairing, $\Delta^{xz}(\mathbf{k})=2\Delta^{xz}(\cos k_x-\cos k_y)$, with $\Delta^{xz}=2\,\mathrm{meV}$, $\Delta^x_{\perp}=16\,\mathrm{meV}$ and $\Delta^z_{\perp}=13\,\mathrm{meV}$. (c) Corresponding tunneling spectra compared with the STM result~\cite{wang2026atomically} (gray region). The line colors and styles shown in (b) apply to all three panels. The parameters in (c) are the same as in (a,b), except that the black dashed curve uses $\Delta^z_{\perp}=11\,\mathrm{meV}$.}\label{fig2}
\end{figure*}

\subsection*{Typical strong-coupling pairing states}
 A natural and constructive strong-coupling viewpoint starts from the interlayer AFM superexchange between the nearly half-filled $d_{z^2}$ orbitals~\cite{Yi_Feng2023,ZhangGM2023DMRG,kaneko2023pair,chen2026evolution}. This superexchange generates intrinsic interlayer $d_{z^2}$-orbital pairing. Despite the low in-plane hopping integral, the $d_{z^2}$-orbital pairs acquire phase coherence through hybridization with the more itinerant $d_{x^2-y^2}$-orbital electrons. Based on this pairing mechanism, the resulting MF Hamiltonian is
\begin{equation}\label{eq:HMFz}
\begin{aligned}
H_{\rm MF}^{(z)}&=H_0^x+H_0^z+H_0^{xz}+H^z_{\rm pair},\\
H^z_{\rm pair}&=\Delta^z_{\perp}\sum_i
\left(c^\dagger_{izt\uparrow}c^\dagger_{izb\downarrow}
-c^\dagger_{izt\downarrow}c^\dagger_{izb\uparrow}+{\rm H.c.}\right).
\end{aligned}
\end{equation}
We project Eq.~(\ref{eq:HMFz}) onto the FSs. The calculated gaps on the $\beta$- and $\gamma$- pockets and the STM spectrum are shown by the blue curves in Fig.~\ref{fig2}(a-c), respectively, where the gray regions denote the experimental results. With $\Delta^z_{\perp}=13\,\mathrm{meV}$, the calculated $\gamma$-pocket gap well fits the measured one. By contrast, the calculated $\beta$-pocket gap vanishes at $\theta=45^\circ$ and remains much smaller than the measured gap away from the BZ diagonal. These diagonal nodes lead to a V-shaped calculated STM spectrum, whereas the STM experiment observes a U-shaped spectrum. Thus, a state with intrinsic pairing restricted to the $d_{z^2}$- orbital sector is challenged by the experimentally observed nodeless gap.

The nodes on the $\alpha/\beta$ pockets arise from the absence of intrinsic $d_{x^2-y^2}$-orbital pairing $H^x_{\rm pair}$ in Eq.~(\ref{eq:HMFz}), since their diagonal gaps are determined solely by $H^x_{\rm pair}$, as shown in Eq.~(\ref{eq:H_MF_diag}). The $d_{x^2-y^2}$ sector therefore retains normal state on the BZ diagonal, producing the $\alpha/\beta$ nodes and the V-shaped STM spectrum. These diagonal nodes survive the inclusion of symmetry-allowed interorbital pairing, as established by the mirror-symmetry argument above. Specifically, as shown in Supplementary Fig.~\ref{fig:interorbital_correction}, a representative nearest-neighbor intralayer interorbital pair field modifies the gaps away from the diagonal, while the diagonal $\alpha/\beta$-pocket node remains unchanged.

Notably, as shown in Supplementary Note~1, orbital hybridization yields a finite real-space interlayer pairing expectation value in the $d_{x^2-y^2}$ sector $\left\langle\Delta_x\right\rangle\equiv\frac{1}{\sqrt{2}}\left\langle c_{ixt\uparrow}c_{ixb\downarrow}-c_{ixt\downarrow}c_{ixb\uparrow}\right\rangle$. Despite its finite real-space value which is a momentum-space sum, the momentum-resolved contribution vanishes on the BZ diagonal, where the hybridization is zero, and therefore cannot remove the $\alpha/\beta$ nodes. However, if an intrinsic $d_{x^2-y^2}$-orbital pairing interaction is present, this finite pairing expectation value allows the pairing interaction Hamiltonian to be decomposed at the mean-field level into an intrinsic, self-consistent pair field term in the $d_{x^2-y^2}$ channel, which can remove the $\alpha/\beta$ nodes as clarified below. Therefore, a substantial intrinsic $d_{x^2-y^2}$-orbital pairing component is indispensable for understanding the nearly isotropic, nodeless gaps observed experimentally.

An appreciable intrinsic $d_{x^2-y^2}$-orbital pair field is also directly indicated by the experimental diagonal gaps. The measured $\beta$- and $\gamma$-pocket gaps near $\theta=45^\circ$ are approximately $16$ and $13$ meV, respectively~\cite{shen2025anomalous}. According to the general symmetry argument above, these diagonal gaps directly probe the intrinsic pairing in the $d_{x^2-y^2}$ and $d_{z^2}$ sectors, respectively.

Accordingly, we extend the MF Hamiltonian in Eq.~(\ref{eq:HMFz}) to include an intrinsic interlayer $d_{x^2-y^2}$-orbital pair field,
\begin{equation}\label{eq:Hxpair}
H^x_{\rm pair}=\Delta^x_{\perp}\sum_i\left(
c^\dagger_{ixt\uparrow}c^\dagger_{ixb\downarrow}
-c^\dagger_{ixt\downarrow}c^\dagger_{ixb\uparrow}
+{\rm H.c.}\right).
\end{equation}

As shown in Fig.~\ref{fig2}(a,b), setting representative $\Delta^x_{\perp}=20\,\mathrm{meV}$, the calculated gap function (red dashed) on the $\beta$-pocket becomes nodeless and comparable with ARPES data (gray region), while agreement with the measured $\gamma$-pocket gap is preserved. The corresponding STM spectrum in Fig.~\ref{fig2}(c) is U-shaped. The pairing state including a $d_{x^2-y^2}$-orbital component therefore qualitatively reproduces the fully gapped ARPES/STM spectra.

Quantitatively, the fit to the ARPES/STM spectra can be further improved by including a weak interorbital pairing term. Interorbital pairing modifies the gap away from the BZ diagonal, as shown above, and a finite component is supported by several theoretical studies~\cite{qiu2025pairing,wu2026interorbital,10.1093/nsr/nwaf353}. For a minimal fit, we consider the strongest symmetry-allowed intraorbital and interorbital pairing channels. In strong-coupling theories that yield $s$-wave pairing, the leading intraorbital component is typically pairing on vertical bonds~\cite{Yi_Feng2023,lu2023bilayertJ,qiu2025pairing}. For the interorbital pairing component, in-plane nearest-neighbor pairing $\Delta^{xz}$ (see Methods) is considered here, since $s$-wave pairing between the two orbitals on vertical bonds is forbidden by symmetry. The $(\Delta^x_{\perp},\Delta^z_{\perp})$ parameter space is scanned by optimizing $\Delta^{xz}$ at each point to minimize the mean squared difference between the calculated and measured ARPES gaps on the $\beta$ and $\gamma$ pockets. As shown in Supplementary Fig.~3, the fitting loss reaches its minimum for comparable intraorbital pairing strengths in the two orbital channels and a small interorbital pairing amplitude $\Delta^{xz}$. With $\Delta^x_{\perp}=16$ meV, $\Delta^z_{\perp}=13$ meV, and $\Delta^{xz}=2$ meV, a parameter set near the minimum of the fitting loss, the gap distributions (black dashed) shown in Figs.~\ref{fig2}(a,b) capture the main features of the ARPES data. Furthermore, using the same $\Delta^{x}_{\perp}$, $\Delta^{xz}$ and slightly different $\Delta^z_{\perp} = 11$ meV, the calculated STM spectrum (black dashed) also agrees well with the U-shaped experimental data, as shown in Fig.~\ref{fig2}(c). The slight variation of $\Delta^z_{\perp}$ may be caused by sample difference which influences the oxygen content and hence the size of the $\gamma$-pocket. The success of these minimal fits to the spectroscopic data therefore further supports the necessity of an appreciable $d_{x^2-y^2}$-orbital pairing component for La$_3$Ni$_2$O$_7$.

Physically, the appreciable intrinsic $d_{x^2-y^2}$-orbital pairing component can be induced by some intrinsic pairing interaction in the $d_{x^2-y^2}$ channel, whose possible origin is analyzed below. At quarter filling, the $d_{x^2-y^2}$-orbital sector lies in the heavily overdoped regime, and its intralayer pairing interaction should be weak. For interlayer $d_{x^2-y^2}$-orbital pairing, the direct interlayer AFM superexchange is too weak to provide the pairing interaction. Instead, Hund's coupling can transfer the strong $d_{z^2}$ interlayer AFM superexchange to the $d_{x^2-y^2}$ electrons, generating an effective interlayer superexchange which serves as the intrinsic pairing interaction in the $d_{x^2-y^2}$-orbital channel~\cite{lu2023bilayertJ,oh2023type2,qu2023bilayer,zhang2023strong,pan2023rno,yang2023strong,wu2024deconfined,Lu2024interplay,PhysRevB.113.174521,shao2025pairing,chen2026unified}. Alternatively, Hund-coupled local $d_{z^2}$ triplons may provide another route to $d_{x^2-y^2}$-orbital pairing~\cite{liu2026triplon}. The observed appreciable $d_{x^2-y^2}$-orbital pairing component is therefore compatible with Hund's-rule-driven pairing mechanisms.

\subsection*{Weak-coupling pairing}
We perform RPA calculations for the two-orbital Hubbard--Kanamori model using different thin-film TB parametrizations available in the literature~\cite{yue2025correlated,Daoxin_Yao2025,cao2025strain,cao2026tunable,li2025substrate,ryee2025optimal} and interaction parameters (see Methods). The leading pairing states obtained in all these calculations have $s$-wave symmetry. Their gap functions consistently exhibit nodes or near-nodes on the $\alpha$ or $\beta$ pocket. Here we present a representative result in Fig.~\ref{fig3}, while the remaining results are shown in Supplementary Fig.~4. In this representative example, the calculated $\beta$-pocket gap develops two nodes near the BZ diagonal, one on each side, in contrast to the nearly isotropic nodeless gap observed by ARPES~\cite{shen2025anomalous,miao2026thermodynamic}. To clarify the origin of these nodes, we Fourier transform the RPA gap function into real-space orbital basis. As summarized in Supplementary Table~\ref{table_gap_real_space_rpa}, two features are prominent. First, the $d_{x^2-y^2}$-orbital pairing component is substantially weaker than its $d_{z^2}$-orbital counterpart. Second, the $d_{x^2-y^2}$-orbital pairing is spatially more extended, possibly associated with stronger in-plane hopping, giving rise to stronger momentum dependence. Its different real-space pairing components then interfere destructively near the BZ diagonal, further suppressing the gap there. Meanwhile, the hybridization-mediated contribution from the $d_{z^2}$-orbital pairing to the $\beta$-pocket gap vanishes on the diagonal. These two effects jointly produce the nodes on the $\beta$ pocket.

In our RPA calculations, the $d_{z^2}$-orbital dominance originates from the flat top of the bonding $d_{z^2}$ band forming the $\gamma$ pocket, whose large density of states enhances pairing in this orbital. A similar $d_{z^2}$ dominance, accompanied by nodes or near-nodes on the $\alpha/\beta$ pockets, appears broadly in previous weak-coupling calculations~\cite{YangF2023,zhang2023structural,zhang2023trends,gao2025robust,yue2025correlated,shao2025band,ryee2025optimal,heier2023competing,PhysRevB.111.104505,ushio2025theoretical,hua2026possible,WangQH2023,HuJP2023,cao2025strain} and is challenged by the fully gapped ARPES/STM spectra.

\begin{figure}[tp]
    \centering
    \includegraphics[width=\columnwidth,keepaspectratio]{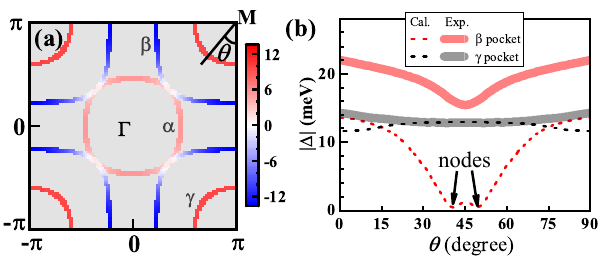}
    \caption{ RPA-calculated superconducting gap. (a) Distribution of the superconducting gap $\Delta(\mathbf{k})$ on the Fermi surfaces. (b) Calculated angle-dependent gap magnitudes on the $\beta$ pocket (red dashed line) and the $\gamma$ pocket (black dashed line), compared with the experimental results (shaded regions). The arrows mark the two nodes at which the calculated $\beta$-pocket gap changes sign. The overall scale of the gap eigenfunction is chosen for comparison with experiment. } \label{fig3}
\end{figure}

\subsection*{Predictions}
 The interlayer $d_{x^2-y^2}$-orbital pair field considered above yields two experimentally testable predictions, as summarized in Fig.~4. First, along the BZ diagonal, the $\alpha$ and $\beta$ states are respectively the interlayer bonding and antibonding combinations of the $d_{x^2-y^2}$ orbitals, $c_{x,\alpha/\beta}=(c_{xt}\pm c_{xb})/\sqrt{2}$. As an interlayer pair field probes the product of the two layer amplitudes, it projects with opposite signs onto the bonding and antibonding sectors, yielding $\Delta_\alpha(\mathbf{k}_{\rm diag})=-\Delta_\beta(\mathbf{k}_{\rm diag})$; see Supplementary Note~5. The two diagonal gaps host equal magnitude but opposite signs. Moreover, because the smaller $\alpha$ pocket spans a narrower region of momentum space, the gap on the $\alpha$ pocket is even more isotropic than that on the $\beta$ pocket, as shown in Fig.~4(a). The predicted magnitude and momentum dependence can be tested by resolving the $\alpha$ pocket with ARPES.

\begin{figure}[tp]
    \centering
    \includegraphics[width=\columnwidth,keepaspectratio]{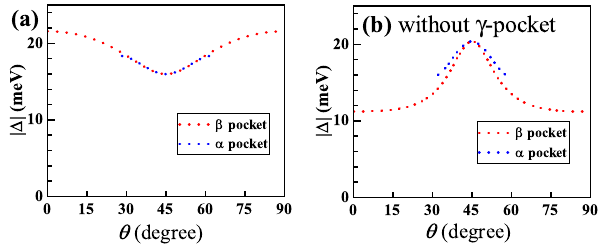}
    \caption{Predicted superconducting gaps on the $\alpha$ and $\beta$ pockets. (a) $\theta$-dependent gap magnitudes calculated using the same pairing parameters as the black dashed curves in Fig.~\ref{fig2}. (b) Corresponding results for the no-$\gamma$-pocket TB model of Ref.~\cite{shao2025pairing}, with only $\Delta^x_{\perp}=20\,\mathrm{meV}$ retained. The red and blue curves denote the $\beta$ and $\alpha$ pockets, respectively, and \mbox{$\theta=45^\circ$} cor\-responds to the BZ diagonal.} \label{fig4}
\end{figure}

Second, when the $\gamma$ pocket is absent, the nearly half-filled $d_{z^2}$ electrons become more localized, suppressing their contribution to coherent pairing. We therefore consider the limiting case $\Delta^z_{\perp}=0$. The interlayer $d_{x^2-y^2}$-orbital pairing state yields full $\alpha/\beta$ gaps that reach their maxima on the BZ diagonal for $\Delta^z_{\perp}=0$, as shown in Fig.~\ref{fig4}(b). These features can be tested by ARPES on films without the $\gamma$ pocket.

\section*{Discussion}
 In conclusion, based on symmetry analysis, we propose that the vanishing of orbital hybridization and interorbital pairing along the BZ diagonal causes orbital-selective pairing on different Fermi pockets, which provides a probe of the orbital character of pairing in La$_3$Ni$_2$O$_7$. Applying this criterion, we find that when intrinsic pairing is present purely in the $d_{z^2}$ orbital, nodes emerge on the $\alpha/\beta$ pockets, a result challenged by the fully gapped ARPES/STM spectra. The additional appreciable intrinsic interlayer $d_{x^2-y^2}$-orbital pairing component removes these nodes, and in combination with a weak symmetry-allowed interorbital pairing term, leads to a minimal fit which well captures the main features of the ARPES/STM spectra. The fit supports an appreciable $d_{x^2-y^2}$-orbital pairing component, which is indispensable for the finite diagonal $\alpha/\beta$ gap in La$_3$Ni$_2$O$_7$. This pairing state is compatible with Hund's-rule-driven pairing mechanisms.

Within this proposed interlayer mechanism, the near-quarter filling of the $d_{x^2-y^2}$ orbital places this pairing sector in the overdoped regime, where $T_c$ is governed primarily by the doping level and the effective interlayer AFM exchange $J^x_{\perp}$. This framework provides a unified account of the available $T_c$-control experiments~\cite{chen2026unified}.

Recently, Scholten \emph{et al.}~\cite{scholten2026nodal} showed that an orbital-antiphase $s_\pm$ state can host accidental nodes on the $\alpha$ or $\beta$ pocket. By contrast, their orbital-in-phase solution yields a nodeless $\beta$-pocket gap, consistent with our fits in Fig.~\ref{fig2} using same-sign $\Delta_\perp^x$ and $\Delta_\perp^z$.

\section*{Methods}
\subsection*{Calculation of the superconducting gap}
For the intraorbital pair fields, we start from the mean-field Hamiltonian
\begin{equation}
\begin{aligned}
H_{\rm MF} &= H_0 + \sum_{m,i}\Delta_{\perp}^{m} \left( c_{imt\uparrow}^\dagger c_{imb\downarrow}^\dagger - c_{imt\downarrow}^\dagger c_{imb\uparrow}^\dagger + \mathrm{H.c.} \right) \\
&=\sum_{\mathbf{k},\sigma}\sum_{mm'\mu\nu}h_{\mu\nu}^{mm'}(\mathbf{k})c_{\mathbf{k}m\mu\sigma}^\dagger c_{\mathbf{k}m'\nu\sigma} \\
&+ \sum_{m\mathbf{k}}\Delta_{\perp}^{m} \left( c_{\mathbf{k}mt\uparrow}^\dagger c_{-\mathbf{k}mb\downarrow}^\dagger - c_{\mathbf{k}mt\downarrow}^\dagger c_{-\mathbf{k}mb\uparrow}^\dagger + \mathrm{H.c.} \right)\notag,
\end{aligned}
\end{equation}
where $m,m'\in\{x,z\}$ label the orbitals and $\mu,\nu\in\{t,b\}$ label the layers. Diagonalizing the matrix $h(\mathbf{k})$ yields the band eigenstates $|\mathbf{k},l\rangle$ with eigenvector components $\xi_{l,m\mu}(\mathbf{k})$ and eigenvalues $\epsilon_l(\mathbf{k})$. 

We reasonably adopt the intraband pairing approximation, in which only the intraband Cooper pairing of the form $c_{\mathbf{k}l\sigma}^\dagger c_{-\mathbf{k}l\sigma'}^\dagger$ is retained. The band-basis mean-field Hamiltonian then reads
\begin{equation}
H_{\rm MF} = \sum_{\mathbf{k},l,\sigma} \epsilon_l(\mathbf{k}) c_{\mathbf{k}l\sigma}^\dagger c_{\mathbf{k}l\sigma}
+ \sum_{\mathbf{k},l} \left[ \Delta_{l}(\mathbf{k}) c_{\mathbf{k}l\uparrow}^\dagger c_{-\mathbf{k}l\downarrow}^\dagger + \mathrm{H.c.} \right]\notag,
\end{equation}
with
\begin{equation}
\Delta_{l}(\mathbf{k}) = 2\Delta_{\perp}^{x}\,
\xi_{l,xt}^*(\mathbf{k}) \, \xi_{l,xb}^*(-\mathbf{k}) + 2\Delta_{\perp}^{z} \, \xi_{l,zt}^*(\mathbf{k}) \, \xi_{l,zb}^*(-\mathbf{k})\notag
\end{equation}

The intraband pairing approximation we adopt is well justified by the normal-state band separations. The only potentially relevant interband channel occurs along the BZ diagonal, where the $\alpha$ and $\beta$ bands approach degeneracy and their energy separation becomes comparable to the superconducting gap. Since both states involved carry $d_{x^2-y^2}$-orbital character, including this channel would not alter our orbital-selective conclusion. Elsewhere, the band separations greatly exceed the superconducting gap, rendering interband pairing negligible.

For the fits including interorbital pairing in Fig.~\ref{fig2}, we project the orbital pairing matrix onto the normal-state bands and retain only its band-diagonal entries.  Here, $\Delta^{xz}(\mathbf{k})=2\Delta^{xz}(\cos k_x-\cos k_y)$, where $\Delta^{xz}$ is the real-space pairing amplitude on an $x$-directed bond, with the opposite sign on a $y$-directed bond.  The gap on each Fermi-surface sheet is given by the magnitude of the corresponding intraband pair field $\Delta_l(\mathbf{k})$.

\subsection*{Calculation of the tunneling spectrum}
At finite temperature, the differential conductance is obtained by convolving the BCS density of states with the thermal kernel $K_T(E)=-\partial f/\partial E$,
\begin{equation}
\frac{dI}{dV}(V) \propto \sum_{l,\mathbf{k}} \left[ u_{l\mathbf{k}}^2 K_T\bigl(E_{l\mathbf{k}}-eV\bigr) + v_{l\mathbf{k}}^2 K_T\bigl(E_{l\mathbf{k}}+eV\bigr) \right]\notag,
\end{equation}
where $u_{l\mathbf{k}}$ and $v_{l\mathbf{k}}$ are the standard BCS coherence factors, $E_{l\mathbf{k}}$ is the quasiparticle dispersion, and
\begin{equation}
K_T(E) = \frac{1}{4k_{\rm B}T}\,\mathrm{sech}^2\!\left(\frac{E}{2k_{\rm B}T}\right)\notag.
\end{equation}

We take $k_{\rm B}T = 0.362$~meV in calculation, corresponding to the experimental measurement temperature. The thermally broadened spectrum is further convolved with a Gaussian of standard deviation $\sigma=2$ meV and normalized to its maximum over the calculated bias interval.

\subsection*{Random-phase approximation calculations}
For the representative RPA calculation in Fig.~\ref{fig3}, we consider the multi-orbital Hubbard-Kanamori Hamiltonian and set the on-site intraorbital Coulomb repulsion strength $U=1\text{ eV}$ and the strength of Hund's coupling $J_H=U/6$, and the on-site interorbital Coulomb repulsion strength $V$ is obtained from the relationship $U=V+2J_H$ \cite{PhysRevB.18.4945}. 
Then we apply the standard RPA approach \cite{takimoto2004strong,yada2005origin,kubo2007pairing,graser2009near,liu2013d+,zhang2022lifshitz,kuroki101unconventional} to obtain the $\mathbf{k}$-space gap function in the full BZ with the particle number per site fixed at $n=1.5$. 
After getting the $\mathbf{k}$-space gap function $\Delta_{l}(\mathbf{k})$ in the full BZ, we obtain the real-space pairing components by Fourier transformation.

\section*{Data availability}
The data supporting the findings of this study are available in the article and its Supplementary Information. Additional data are available from the corresponding author upon reasonable request.
\section*{Code availability}
The codes used in this study are available from the corresponding author upon reasonable request.
\section*{Acknowledgements}
This work was supported by the National Natural Science Foundation of China (Grant Nos.~12574141, 12234016, 12304180, and 12504219), the CAS Superconducting Research Project (Grant No.~SCZX-0101), and the Postdoctoral Fellowship Program and the China Postdoctoral Science Foundation (Grant No.~BX20250175).
\section*{Author contributions}
Y.-B.L. and Z.-Y.S. carried out the calculations. Z.P. and C.L. contributed to the scientific discussions. F.Y. conceived the study and led the writing of the manuscript.
\section*{Competing interests}
The authors declare no competing interests.

\clearpage
\onecolumngrid
\begin{center}\large\bfseries Supplementary Information\end{center}

\setcounter{figure}{0}
\setcounter{table}{0}
\setcounter{equation}{0}
\renewcommand{\figurename}{Supplementary Figure}
\renewcommand{\tablename}{Supplementary Table}
\renewcommand{\theequation}{S\arabic{equation}}
\section*{Supplementary Note 1: Proximity-induced $d_{x^2-y^2}$-orbital pairing expectation}

\begin{figure}[htbp]
    \centering
    \includegraphics[width=.38\linewidth]{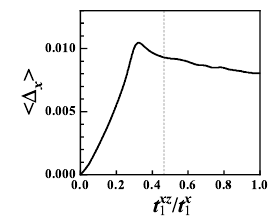}
    \caption{Interlayer $d_{x^2-y^2}$-orbital pairing expectation $\langle\Delta_x\rangle$ as a function of the nearest-neighbor interorbital hopping $t_1^{xz}$, normalized by the intraorbital hopping $t_1^x$. The vertical dashed line marks the hopping ratio used in the main-text calculations.}
    \label{fig:proximity_expectation}
\end{figure}

To verify the proximity-induced pairing expectation discussed in the main text, we evaluate the real-space interlayer anomalous expectation in the $d_{x^2-y^2}$ orbital for the mean-field Hamiltonian in Eq.~(\ref{eq:HMFz}),
\begin{equation}
\left\langle\Delta_x\right\rangle\equiv\frac{1}{\sqrt{2}}
\left\langle c_{ixt\uparrow}c_{ixb\downarrow}
-c_{ixt\downarrow}c_{ixb\uparrow}\right\rangle.
\end{equation}
Although Eq.~(\ref{eq:HMFz}) contains no intrinsic $d_{x^2-y^2}$-orbital pairing term, this expectation can become finite through hybridization with the paired $d_{z^2}$ orbital. We vary the nearest-neighbor interorbital hopping $t_1^{xz}$ while keeping the other parameters fixed. As shown in Fig.~\ref{fig:proximity_expectation}, $\langle\Delta_x\rangle$ vanishes when $t_1^{xz}=0$, for which the two orbital sectors are decoupled, and becomes finite once the hybridization is introduced. This result directly confirms the proximity-effect origin of the induced $d_{x^2-y^2}$-orbital pairing expectation.

The finite real-space expectation is a sum over the full BZ and does not imply a nonzero induced component at every momentum. In particular, its momentum-resolved contribution vanishes on the BZ diagonal together with $H_0^{xz}(\mathbf{k}_{\rm diag})$. The proximity-induced expectation therefore cannot remove the $\alpha/\beta$-pocket nodes and is distinct from an intrinsic $d_{x^2-y^2}$-orbital pairing component.

\section*{Supplementary Note 2: Effect of interorbital pairing on the superconducting gap}
\label{app:interorbital}

\begin{figure}[htbp]
\centering
\includegraphics[width=.68\linewidth]{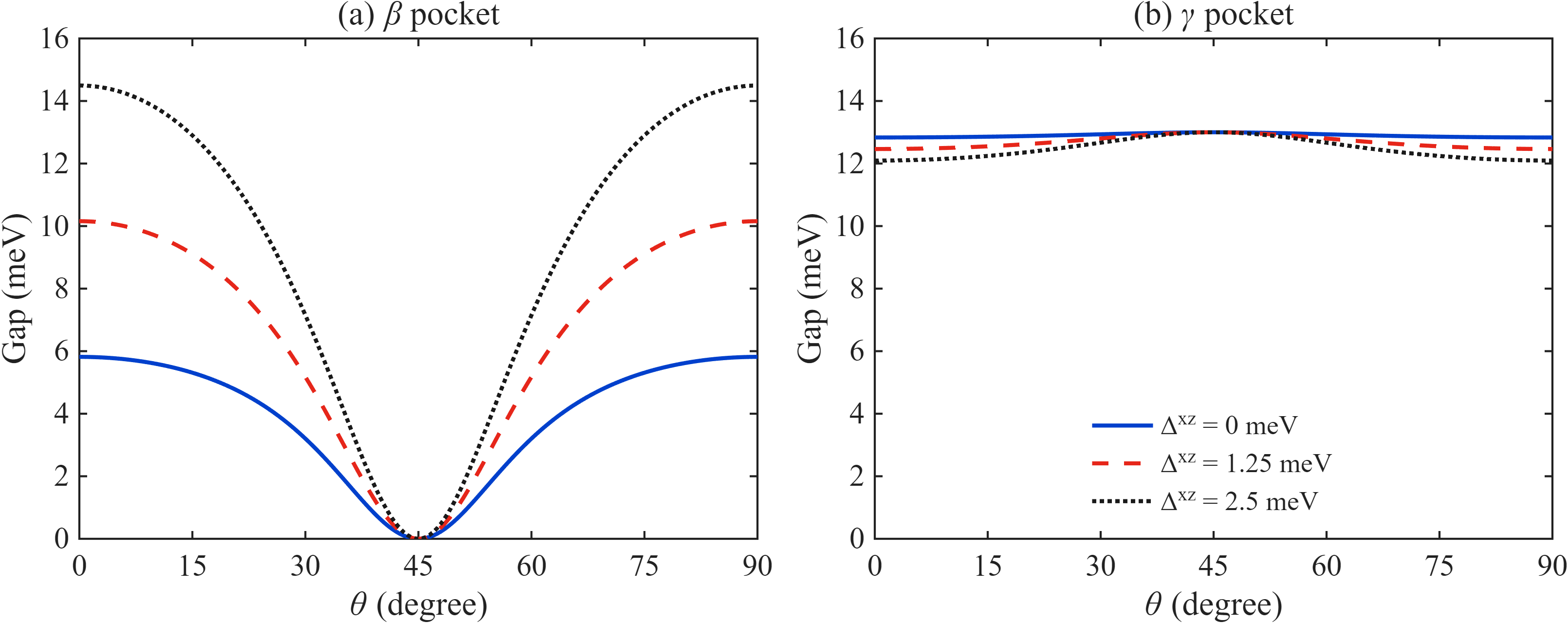}
\caption{Intraband interorbital correction to the restricted $d_{z^2}$ pair-field ansatz. Angular gaps on the $\beta$ and $\gamma$ pockets for intralayer $\Delta^{xz}(\mathbf{k})=2\Delta^{xz}(\cos k_x-\cos k_y)$, with $\Delta^{xz}=0,1.25,2.5$ meV at $\Delta^x_{\perp}=0$ and $\Delta^z_{\perp}=13$ meV. The diagonal $\beta$ node at $45^\circ$ is unchanged.}\label{fig:interorbital_correction}
\end{figure}

The nodes on the BZ diagonal also survive when interorbital pairing, $\Delta^{xz}(\mathbf{k})=2\Delta^{xz}(\cos k_x-\cos k_y)$, is added to the Hamiltonian in Eq.~(\ref{eq:HMFz}) with no intrinsic $d_{x^2-y^2}$-orbital pair field. Supplementary Fig.~\ref{fig:interorbital_correction} presents the quasiparticle gaps obtained within the intraband pairing approximation for $\Delta^{xz}=0,1.25,2.5$ meV, with $\Delta^z_{\perp}=13$ meV. Increasing $\Delta^{xz}$ enhances the $\beta$-pocket gap away from the diagonal and has little effect on the $\gamma$-pocket gap, while the diagonal $\beta$-pocket node remains unchanged. This is because $\Delta^{xz}(\mathbf{k}_{\rm diag})=0$ and the orbital hybridization vanishes on the diagonal, leading to the survival of the $\beta$-pocket node in the absence of an intrinsic $d_{x^2-y^2}$-orbital pair field.

\section*{Supplementary Note 3: Fitting the ARPES gaps with pair fields}
\label{app:arpes_profile}

The main text presents a set of pairing parameters that captures the main features of the experimental gaps. Here we scan the intraorbital pairing amplitudes $\Delta^x_{\perp}$ and $\Delta^z_{\perp}$ to fit the ARPES gaps on the $\beta$ and $\gamma$ pockets.

The fit uses the original experimental data for sample 3 in Figs.~2D,E of Ref.~\cite{sm-shen2025anomalous}, comprising gap measurements at four different angles on each of the $\beta$ and $\gamma$ pockets. For each pair of intraorbital fields $\Delta^x_{\perp}$ and $\Delta^z_{\perp}$, the real-space interorbital bond amplitude $\Delta^{xz}$ is optimized independently, yielding the fitting loss
{
\begin{equation}
\mathcal{L}(\Delta^x_{\perp},\Delta^z_{\perp})
=\min_{\Delta^{xz}}\frac{1}{8}\sum_{i=1}^{8}
\left[|\Delta^{\mathrm{calc}}_i(\Delta^x_{\perp},\Delta^z_{\perp},\Delta^{xz})|-\Delta_i^{\mathrm{exp}}\right]^2,
\label{eq:arpes_profile_loss}
\end{equation}}
where $\Delta_i^{\mathrm{exp}}$ is the measured gap of the $i$th data point and $\Delta_i^{\mathrm{calc}}$ is the corresponding calculated gap. The calculations use the TB model of Ref.~\cite{sm-Daoxin_Yao2025}, as in Fig.~\ref{fig2} of the main text.

\begin{figure}[htbp]
\centering
\includegraphics[width=.90\linewidth]{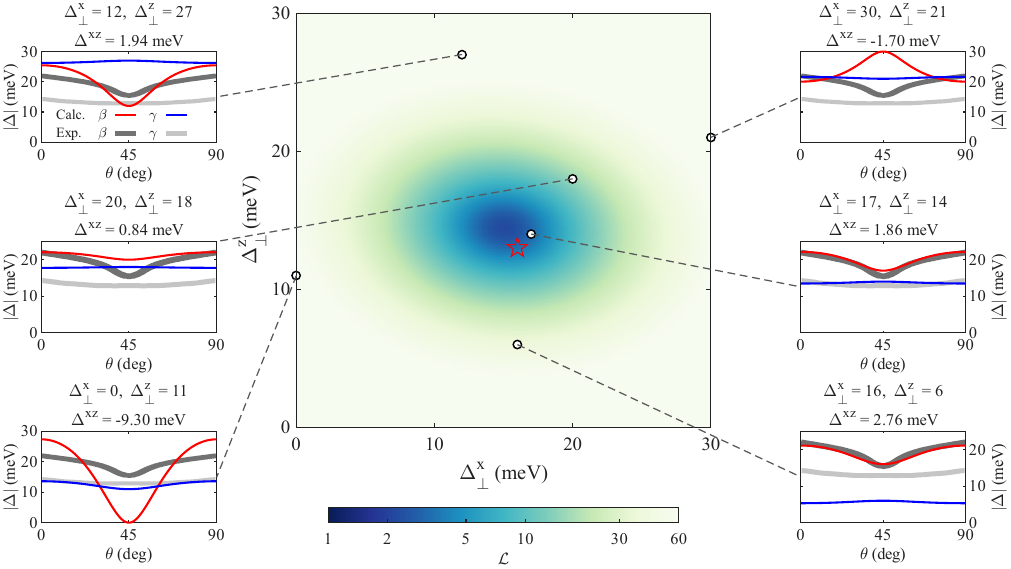}
\caption{ARPES fitting loss. The color map shows the mean squared fitting loss $\mathcal{L}$ for the original sample-3 gap data in Figs.~2D,E of Ref.~\cite{sm-shen2025anomalous}, using equal weights and optimizing $\Delta^{xz}$ independently for each pair of $\Delta^x_{\perp}$ and $\Delta^z_{\perp}$. The color scale is logarithmic. The red star marks the main-text parameters $(\Delta^x_{\perp},\Delta^z_{\perp})=(16,13)$ meV, with $\Delta^{xz}=2$ meV. Open circles mark the parameter sets shown in the insets, with dashed lines connecting each point to its corresponding inset. Red and blue curves show the calculated $\beta$- and $\gamma$-pocket gap magnitudes as functions of $\theta$, respectively; the three pairing amplitudes are listed above each inset in meV. The dark and light gray bands reproduce the experimental guides to the eye for the $\beta$ and $\gamma$ pockets, respectively.}
\label{fig:arpes_profile}
\end{figure}

Supplementary Fig.~\ref{fig:arpes_profile} shows the resulting fitting loss as a function of $\Delta^x_{\perp}$ and $\Delta^z_{\perp}$. The main-text parameters $(\Delta^x_{\perp},\Delta^z_{\perp},\Delta^{xz})=(16,13,2)$ meV, marked by the red star, lie close to the best-fit point. The substantially larger fitting loss for small $\Delta^x_{\perp}$, even after optimizing $\Delta^{xz}$, further supports the indispensable role of the $d_{x^2-y^2}$-orbital pairing component in reproducing the experimental gaps.

\section*{Supplementary Note 4: Random Phase Approximation Study}
The ratios between some dominant pairing components and the interlayer pairing of the $d_{z^2}$ orbital $\Delta^{z,\perp}_{(0,0)}$ are shown in Table \ref{table_gap_real_space_rpa}. For each case in $d_{z^2}$ intralayer pairing, $d_{z^2}$ interlayer pairing, $d_{x^2-y^2}$ intralayer pairing, $d_{x^2-y^2}$ interlayer pairing and interorbital pairing, only the relatively large values are shown.

\begin{table}[htbp]
    \centering
    \caption{The ratios between some dominant pairing components and the interlayer pairing of the $d_{z^2}$ orbital obtained from the RPA study. In the superscripts, $z/x$ represents the pairing of the $d_{z^2}$ orbital or the $d_{x^2-y^2}$ orbital, 
    and $\parallel$ and $\perp$ represent intralayer pairing and interlayer pairing, respectively. $(r_x,r_y)$ in the subscript represents the pairing between site $\mathbf{R}$ and site $\mathbf{R}+r_x\mathbf{e}_x+r_y\mathbf{e}_y$. 
    Only the values with magnitudes $\gtrsim 0.1$ are listed here. } \label{table_gap_real_space_rpa} 
    \begin{tabular}{cc}
        \hline
        Pairing Component & Value \\
        \hline
        $\Delta^{z,\parallel}_{(0,0)}$ & $-0.381$ \\
        $\Delta^{z,\perp}_{(0,0)}$ & $1.0$ \\
        $\Delta^{x,\parallel}_{(1,1)}$ & $0.095$ \\
        $\Delta^{x,\perp}_{(0,0)}$ & $0.457$ \\
        $\Delta^{x,\perp}_{(1,1)}$ & $-0.125$ \\

        \hline
    \end{tabular}
\end{table}

\begin{figure}[htbp]
    \centering
    \includegraphics[width=\textwidth,height=0.74\textheight,keepaspectratio]{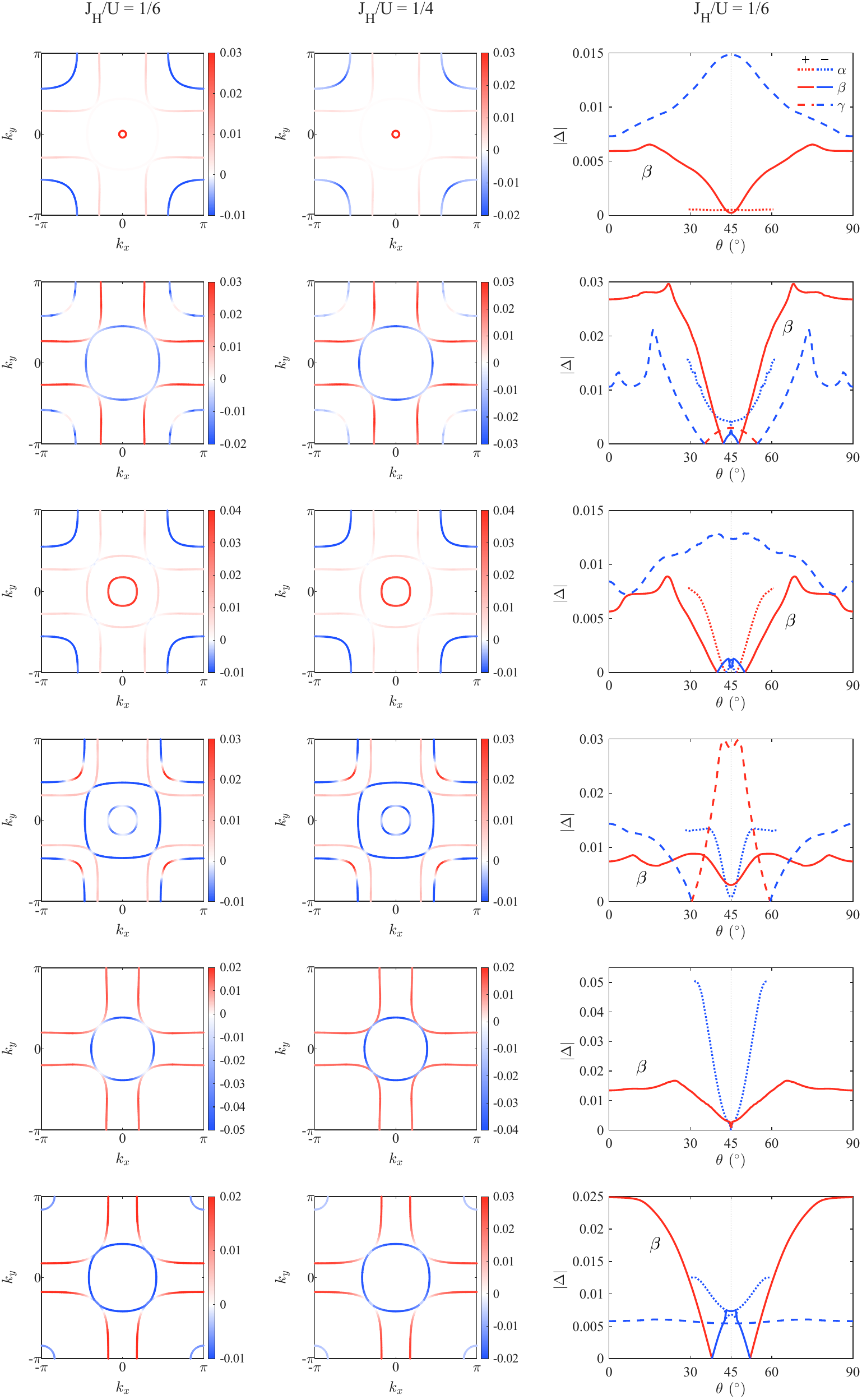}
    \caption{Leading $s$-wave gap eigenfunctions for six thin-film TB models. The first and second columns show the Fermi-surface distributions for $J_H/U=1/6$ and $1/4$, respectively. White in the Fermi-surface maps denotes zero. The third column shows $|\Delta|$ for $J_H/U=1/6$ as a function of the angle $\theta$ defined in the main text. Solid, dotted, and dashed curves denote the $\beta$, $\alpha$, and $\gamma$ pockets, respectively; red and blue indicate positive and negative gap values, respectively. From top to bottom: (1) The film parameters in Table 2 of Ref.~\cite{sm-yue2025correlated}, with $U=0.84$ eV. (2) The half-unit-cell parameters in Table 3 of Ref.~\cite{sm-Daoxin_Yao2025}, with $U=1.08$ and $1.02$ eV in the first and second columns, respectively. (3) The parameters for $2\%$ in-plane compression and zero $c$-axis expansion in Table II of Ref.~\cite{sm-cao2025strain}, with $U=0.60$ eV. (4) The $d_{\mathrm{Ni-Ni}}=4.05$ \AA{} parameters in Table S2 of Ref.~\cite{sm-cao2026tunable}, with $U=0.84$ eV. (5) The first-bilayer parameters in Table 2 of Ref.~\cite{sm-li2025substrate}, with the intralayer parameters averaged over layers 1 and 2, at $U=1.02$ eV. (6) The $-2\%$ strain parameters in Table S1 of Ref.~\cite{sm-ryee2025optimal}, with $U=1.02$ eV.}
    \label{fig:rpa_thinfilm}
\end{figure}

Our RPA results, together with previous weak-coupling calculations, indicate that nodes or near-nodes on the $\alpha$ or $\beta$ pocket are a general feature of weak-coupling gap structures, rather than a consequence of a particular TB model or interaction parameter set. As shown in Fig.~\ref{fig:rpa_thinfilm}, additional RPA calculations using six thin-film TB models~\cite{sm-yue2025correlated,sm-Daoxin_Yao2025,sm-cao2025strain,sm-cao2026tunable,sm-li2025substrate,sm-ryee2025optimal} and different interaction parameters further support this conclusion. For all six TB models, the filling is fixed at $n=1.5$ electrons per Ni site, and hopping terms are retained up to third-nearest in-plane neighbors. The leading pairing states in all cases shown in Supplementary Fig.~\ref{fig:rpa_thinfilm} have $s$-wave symmetry and exhibit nodes or near-nodes near the BZ diagonal on the $\alpha$ or $\beta$ pocket, in contrast to the fully gapped ARPES/STM spectra.

\section*{Supplementary Note 5: Opposite signs of the $\alpha$- and $\beta$-pocket gaps}

Here we derive the relative sign between the $\alpha$- and $\beta$-pocket gaps along the BZ diagonal. Because the orbital hybridization vanishes there, the $\alpha$ and $\beta$ states have pure $d_{x^2-y^2}$-orbital character and are respectively the bonding and antibonding combinations of the two layers. Their operators are related by
\begin{equation}
\begin{pmatrix}
c_{\mathbf{k}\alpha\sigma}\\
c_{\mathbf{k}\beta\sigma}
\end{pmatrix}
=U^\dagger
\begin{pmatrix}
c_{\mathbf{k}tx\sigma}\\
c_{\mathbf{k}bx\sigma}
\end{pmatrix},
\qquad
U=\frac{1}{\sqrt{2}}
\begin{pmatrix}
1&1\\
1&-1
\end{pmatrix}.
\end{equation}

The interlayer $d_{x^2-y^2}$-orbital pair field has amplitude $\Delta^x_{\perp}$. In the layer basis $(t,b)$, its pairing matrix is purely off-diagonal,
\begin{equation}
\hat{\Delta}^{x}_{\rm layer}(\mathbf{k})
=\Delta^x_{\perp}
\begin{pmatrix}
0&1\\
1&0
\end{pmatrix}
=\Delta^x_{\perp}\tau_x.
\end{equation}
Transforming this matrix to the bonding--antibonding basis gives
\begin{equation}
\begin{aligned}
\hat{\Delta}^{x}_{\rm band}(\mathbf{k})
&=U^\dagger\hat{\Delta}^{x}_{\rm layer}(\mathbf{k})U^*\\
&=\Delta^x_{\perp}
\begin{pmatrix}
1&0\\
0&-1
\end{pmatrix}.
\end{aligned}
\end{equation}
Therefore, on the BZ diagonal,
\begin{equation}
\Delta_\alpha(\mathbf{k}_{\rm diag})
=\Delta^x_{\perp}
=-\Delta_\beta(\mathbf{k}_{\rm diag}).
\end{equation}
The overall sign depends on the common gauge convention, whereas the relative minus sign is gauge invariant. The interorbital pair field vanishes on the BZ diagonal and therefore does not alter this relative sign.

\makeatletter
\let\@FMN@list\@empty
\patchcmd{\NAT@bibsetnum}{LastBibItem}{LastSupplementaryBibItem}{}{}
\patchcmd{\endthebibliography}{LastBibItem}{LastSupplementaryBibItem}{}{}
\makeatother
\renewcommand{\refname}{Supplementary References}

\begin{thebibliography}{81}
\expandafter\ifx\csname url\endcsname\relax
  \def\url#1{\texttt{#1}}\fi
\expandafter\ifx\csname urlprefix\endcsname\relax\def\urlprefix{URL }\fi
\providecommand{\bibinfo}[2]{#2}
\providecommand{\eprint}[2][]{\url{#2}}

\bibitem{Wang2023LNO}
\bibinfo{author}{Sun, H.} \emph{et~al.}
\newblock \bibinfo{title}{Signatures of superconductivity near 80~{K} in a
  nickelate under high pressure}.
\newblock \emph{\bibinfo{journal}{Nature}} \textbf{\bibinfo{volume}{621}},
  \bibinfo{pages}{493--498} (\bibinfo{year}{2023}).

\bibitem{YuanHQ2023LNO}
\bibinfo{author}{Zhang, Y.} \emph{et~al.}
\newblock \bibinfo{title}{High-temperature superconductivity with zero
  resistance and strange-metal behaviour in {L}a$_3${N}i$_2${O}$_{7-\delta}$}.
\newblock \emph{\bibinfo{journal}{Nat. Phys.}}
  \textbf{\bibinfo{volume}{20}}, \bibinfo{pages}{1269--1273}
  (\bibinfo{year}{2024}).

\bibitem{Wang2023LNOb}
\bibinfo{author}{Hou, J.} \emph{et~al.}
\newblock \bibinfo{title}{Emergence of high-temperature superconducting phase
  in pressurized {L}a$_{3}${N}i$_{2}${O}$_7$ crystals}.
\newblock \emph{\bibinfo{journal}{Chin. Phys. Lett.}}
  \textbf{\bibinfo{volume}{40}}, \bibinfo{pages}{117302}
  (\bibinfo{year}{2023}).

\bibitem{wang2023la2prnio7}
\bibinfo{author}{Wang, G.} \emph{et~al.}
\newblock \bibinfo{title}{Observation of high-temperature superconductivity in
  the high-pressure tetragonal phase of {L}a$_2${P}r{N}i$_2${O}$_{7-\delta}$}.
\newblock Preprint at \url{https://arxiv.org/abs/2311.08212} (2023).

\bibitem{liu2025superconductivity}
\bibinfo{author}{Liu, Y.} \emph{et~al.}
\newblock \bibinfo{title}{Superconductivity and normal-state transport in
  compressively strained {L}a$_2${P}r{N}i$_2${O}$_7$ thin films}.
\newblock \emph{\bibinfo{journal}{Nat. Mater.}}
  \textbf{\bibinfo{volume}{24}}, \bibinfo{pages}{1221--1227}
  (\bibinfo{year}{2025}).

\bibitem{wang2023structure}
\bibinfo{author}{Wang, L.} \emph{et~al.}
\newblock \bibinfo{title}{Structure responsible for the superconducting state
  in {L}a$_3${N}i$_2${O}$_7$ at high-pressure and low-temperature conditions}.
\newblock \emph{\bibinfo{journal}{J. Am. Chem. Soc.}}
  \textbf{\bibinfo{volume}{146}}, \bibinfo{pages}{7506--7514}
  (\bibinfo{year}{2024}).

\bibitem{geisler2023structural}
\bibinfo{author}{Geisler, B.}, \bibinfo{author}{Hamlin, J.~J.},
  \bibinfo{author}{Stewart, G.~R.}, \bibinfo{author}{Hennig, R.~G.} \&
  \bibinfo{author}{Hirschfeld, P.~J.}
\newblock \bibinfo{title}{Structural transitions, octahedral rotations, and
  electronic properties of ${A}_3${N}i$_2${O}$_7$ rare-earth nickelates under
  high pressure}.
\newblock \emph{\bibinfo{journal}{npj Quantum Mater.}}
  \textbf{\bibinfo{volume}{9}}, \bibinfo{pages}{38} (\bibinfo{year}{2024}).

\bibitem{bhatt2025resolving}
\bibinfo{author}{Bhatt, L.} \emph{et~al.}
\newblock \bibinfo{title}{Structural modifications in strain-engineered bilayer nickelate thin films}.
\newblock \emph{\bibinfo{journal}{Nature}} \textbf{\bibinfo{volume}{653}}, \bibinfo{pages}{76--82} (\bibinfo{year}{2026}).

\bibitem{PhysRevB.112.L140504}
\bibinfo{author}{Yi, X.-W.}, \bibinfo{author}{Li, W.}, \bibinfo{author}{You,
  J.-Y.}, \bibinfo{author}{Gu, B.} \& \bibinfo{author}{Su, G.}
\newblock \bibinfo{title}{Unifying strain- and pressure-driven
  superconductivity in {L}a$_3${N}i$_2${O}$_7$: Suppressed charge and spin
  density waves and enhanced interlayer coupling}.
\newblock \emph{\bibinfo{journal}{Phys. Rev. B}}
  \textbf{\bibinfo{volume}{112}}, \bibinfo{pages}{L140504}
  (\bibinfo{year}{2025}).

\bibitem{Li2024ele}
\bibinfo{author}{Li, Y.} \emph{et~al.}
\newblock \bibinfo{title}{Electronic correlation and pseudogap-like behavior of
  high-temperature superconductor {L}a$_3${N}i$_2${O}$_7$}.
\newblock \emph{\bibinfo{journal}{Chin. Phys. Lett.}}
  \textbf{\bibinfo{volume}{41}}, \bibinfo{pages}{087402}
  (\bibinfo{year}{2024}).

\bibitem{yang2024orbital}
\bibinfo{author}{Yang, J.} \emph{et~al.}
\newblock \bibinfo{title}{Orbital-dependent electron correlation in
  double-layer nickelate {L}a$_3${N}i$_2${O}$_7$}.
\newblock \emph{\bibinfo{journal}{Nat. Commun.}} \textbf{\bibinfo{volume}{15}},
  \bibinfo{pages}{4373} (\bibinfo{year}{2024}).

\bibitem{chen2024electronic}
\bibinfo{author}{Chen, X.} \emph{et~al.}
\newblock \bibinfo{title}{Electronic and magnetic excitations in
  {L}a$_3${N}i$_2${O}$_7$}.
\newblock \emph{\bibinfo{journal}{Nat. Commun.}} \textbf{\bibinfo{volume}{15}},
  \bibinfo{pages}{9597} (\bibinfo{year}{2024}).

\bibitem{YaoDX2023}
\bibinfo{author}{Luo, Z.}, \bibinfo{author}{Hu, X.}, \bibinfo{author}{Wang,
  M.}, \bibinfo{author}{W\'u, W.} \& \bibinfo{author}{Yao, D.-X.}
\newblock \bibinfo{title}{Bilayer two-orbital model of {L}a$_3${N}i$_2${O}$_7$
  under pressure}.
\newblock \emph{\bibinfo{journal}{Phys. Rev. Lett.}}
  \textbf{\bibinfo{volume}{131}}, \bibinfo{pages}{126001}
  (\bibinfo{year}{2023}).

\bibitem{Yubo_Liu2024}
\bibinfo{author}{Liu, Y.-B.} \emph{et~al.}
\newblock \bibinfo{title}{Origin of the diagonal double-stripe
  spin density wave and potential superconductivity in bulk
  {L}a$_3${N}i$_2${O}$_7$ at ambient pressure}.
\newblock \emph{\bibinfo{journal}{Phys. Rev. B}}
  \textbf{\bibinfo{volume}{112}}, \bibinfo{pages}{014510}
  (\bibinfo{year}{2025}).

\bibitem{li2025photoemission}
\bibinfo{author}{Li, P.} \emph{et~al.}
\newblock \bibinfo{title}{Angle-resolved photoemission spectroscopy of
  superconducting ({L}a,{P}r)$_3${N}i$_2${O}$_7$/{S}r{L}a{A}l{O}$_4$ heterostructures}.
\newblock \emph{\bibinfo{journal}{Natl. Sci. Rev.}}
  \textbf{\bibinfo{volume}{12}}, \bibinfo{pages}{nwaf205}
  (\bibinfo{year}{2025}).

\bibitem{wang2025electronic}
\bibinfo{author}{Wang, B.~Y.} \emph{et~al.}
\newblock \bibinfo{title}{Electronic structure of compressively strained bilayer nickelate thin film}.
\newblock \emph{\bibinfo{journal}{Phys. Rev. X}} \textbf{\bibinfo{volume}{16}}, \bibinfo{pages}{031008} (\bibinfo{year}{2026}).

\bibitem{sun2025observation}
\bibinfo{author}{Sun, W.} \emph{et~al.}
\newblock \bibinfo{title}{Observation of superconductivity-induced leading-edge
  gap in a bilayer nickelate}.
\newblock \emph{\bibinfo{journal}{Nat. Phys.}} \textbf{\bibinfo{volume}{22}}, \bibinfo{pages}{1409--1415}
  (\bibinfo{year}{2026}).

\bibitem{yue2025correlated}
\bibinfo{author}{Yue, C.} \emph{et~al.}
\newblock \bibinfo{title}{Correlated electronic structures and unconventional
  superconductivity in bilayer nickelate heterostructures}.
\newblock \emph{\bibinfo{journal}{Natl. Sci. Rev.}}
  \textbf{\bibinfo{volume}{12}}, \bibinfo{pages}{nwaf253}
  (\bibinfo{year}{2025}).

\bibitem{shao2025band}
\bibinfo{author}{Shao, Z.-Y.}, \bibinfo{author}{Liu, Y.-B.},
  \bibinfo{author}{Liu, M.} \& \bibinfo{author}{Yang, F.}
\newblock \bibinfo{title}{Band structure and pairing nature of
  {L}a$_3${N}i$_2${O}$_7$ thin film at ambient pressure}.
\newblock \emph{\bibinfo{journal}{Phys. Rev. B}}
  \textbf{\bibinfo{volume}{112}}, \bibinfo{pages}{024506}
  (\bibinfo{year}{2025}).

\bibitem{Daoxin_Yao2025}
\bibinfo{author}{Hu, X.}, \bibinfo{author}{Qiu, W.}, \bibinfo{author}{Chen,
  C.-Q.}, \bibinfo{author}{Luo, Z.} \& \bibinfo{author}{Yao, D.-X.}
\newblock \bibinfo{title}{Electronic structures and multi-orbital models of
  {L}a$_3${N}i$_2${O}$_7$ thin films at ambient pressure}.
\newblock \emph{\bibinfo{journal}{Commun. Phys.}}
  \textbf{\bibinfo{volume}{8}}, \bibinfo{pages}{506} (\bibinfo{year}{2025}).

\bibitem{ushio2025theoretical}
\bibinfo{author}{Ushio, K.} \emph{et~al.}
\newblock \bibinfo{title}{Theoretical study on ambient-pressure
  superconductivity in {L}a$_3${N}i$_2${O}$_7$ thin films: Structural analysis,
  model construction, and robustness of ${s}^{\pm}$-wave pairing}.
\newblock \emph{\bibinfo{journal}{Phys. Rev. B}}
  \textbf{\bibinfo{volume}{114}}, \bibinfo{pages}{014501}
  (\bibinfo{year}{2026}).

\bibitem{cao2025strain}
\bibinfo{author}{Cao, Y.-H.}, \bibinfo{author}{Jiang, K.-Y.},
  \bibinfo{author}{Lu, H.-Y.}, \bibinfo{author}{Wang, D.} \&
  \bibinfo{author}{Wang, Q.-H.}
\newblock \bibinfo{title}{Strain-engineered electronic structure and
  superconductivity in {L}a$_3${N}i$_2${O}$_7$ thin films}.
\newblock \emph{\bibinfo{journal}{Sci. China Phys. Mech. Astron.}}
  \textbf{\bibinfo{volume}{69}}, \bibinfo{pages}{247412}
  (\bibinfo{year}{2026}).

\bibitem{shao2025pairing}
\bibinfo{author}{Shao, Z.-Y.} \emph{et~al.}
\newblock \bibinfo{title}{Pairing without $\gamma$-pocket in the
  {L}a$_3${N}i$_2${O}$_7$ thin film}.
\newblock \emph{\bibinfo{journal}{Commun. Phys.}} (2026). DOI: 10.1038/s42005-026-02865-4.

\bibitem{shi2025effect}
\bibinfo{author}{Shi, H.} \emph{et~al.}
\newblock \bibinfo{title}{The effect of carrier doping and thickness on the
  electronic structures of {L}a$_3${N}i$_2${O}$_7$ thin films}.
\newblock \emph{\bibinfo{journal}{Chin. Phys. Lett.}}
  \textbf{\bibinfo{volume}{42}}, \bibinfo{pages}{080708}
  (\bibinfo{year}{2025}).

\bibitem{bheemavarapu2025strain}
\bibinfo{author}{Bheemavarapu, S.}
\newblock \bibinfo{title}{Strain-tuned structural, electronic, and
  superconducting properties of thin-film {L}a$_3${N}i$_2${O}$_7$}.
\newblock Preprint at \url{https://arxiv.org/abs/2512.23630} (2025).

\bibitem{wang2025origin}
\bibinfo{author}{Wang, H.-X.} \emph{et~al.}
\newblock \bibinfo{title}{Spin stripes and superconductivity in bilayer nickelates}.
\newblock \emph{\bibinfo{journal}{Phys. Rev. Lett.}} \textbf{\bibinfo{volume}{137}}, \bibinfo{pages}{086502} (\bibinfo{year}{2026}).

\bibitem{yang2025magnetism}
\bibinfo{author}{Yang, H.} \& \bibinfo{author}{Zhang, Y.-H.}
\newblock \bibinfo{title}{Magnetism and superconductivity in bilayer
  nickelate}.
\newblock \emph{\bibinfo{journal}{Phys. Rev. B}} \textbf{\bibinfo{volume}{114}}, \bibinfo{pages}{L171104} (\bibinfo{year}{2026}).

\bibitem{shen2025anomalous}
\bibinfo{author}{Shen, J.} \emph{et~al.}
\newblock \bibinfo{title}{Nodeless superconducting gap and electron-boson
  coupling in ({L}a,{P}r,{S}m)$_3${N}i$_2${O}$_7$ films}.
\newblock \emph{\bibinfo{journal}{Science}} \textbf{\bibinfo{volume}{392}},
  \bibinfo{pages}{1396--1400} (\bibinfo{year}{2026}).

\bibitem{fan2025superconducting}
\bibinfo{author}{Fan, S.} \emph{et~al.}
\newblock \bibinfo{title}{Single-particle tunneling spectrum with a robust
  superconducting gap in {L}a$_2${P}r{N}i$_2${O}$_7$ thin films at ambient
  pressure}.
\newblock \emph{\bibinfo{journal}{Sci. Adv.}}
  \textbf{\bibinfo{volume}{12}}, \bibinfo{pages}{eaeg2429}
  (\bibinfo{year}{2026}).

\bibitem{wang2026atomically}
\bibinfo{author}{Wang, X.} \emph{et~al.}
\newblock \bibinfo{title}{Atomically resolved intrinsic superconducting gap in
  ({L}a,{P}r)$_3${N}i$_2${O}$_7$ films}.
\newblock Preprint at \url{https://arxiv.org/abs/2605.14806} (2026).

\bibitem{liang2026observation}
\bibinfo{author}{Liang, Z.} \emph{et~al.}
\newblock \bibinfo{title}{Observation of flat-bottom {U}-shaped energy gap in
  high-{$T_c$} nickelate ({L}a,{P}r)$_3${N}i$_2${O}$_7$ thin films}.
\newblock Preprint at \url{https://arxiv.org/abs/2605.15703} (2026).

\bibitem{miao2026thermodynamic}
\bibinfo{author}{Miao, Y.} \emph{et~al.}
\newblock \bibinfo{title}{Thermodynamic phase transition, pairing symmetry and
  {F}ermi surface topology in {R}uddlesden--{P}opper nickelate films}.
\newblock Preprint at \url{https://arxiv.org/abs/2608.03908} (2026).

\bibitem{li2025enhanced}
\bibinfo{author}{Li, Q.} \emph{et~al.}
\newblock \bibinfo{title}{Enhanced superconductivity in the compressively
  strained bilayer nickelate thin films by pressure}.
\newblock \emph{\bibinfo{journal}{Nat. Commun.}} \textbf{\bibinfo{volume}{17}},
  \bibinfo{pages}{3276} (\bibinfo{year}{2026}).

\bibitem{10.1093/nsr/nwag151}
\bibinfo{author}{Zhou, G.} \emph{et~al.}
\newblock \bibinfo{title}{Superconductivity onset above 60~{K} in
  ambient-pressure nickelate films}.
\newblock \emph{\bibinfo{journal}{Natl. Sci. Rev.}}
  \textbf{\bibinfo{volume}{13}}, \bibinfo{pages}{nwag151}
  (\bibinfo{year}{2026}).

\bibitem{YangF2023}
\bibinfo{author}{Liu, Y.-B.}, \bibinfo{author}{Mei, J.-W.},
  \bibinfo{author}{Ye, F.}, \bibinfo{author}{Chen, W.-Q.} \&
  \bibinfo{author}{Yang, F.}
\newblock \bibinfo{title}{s$^{\pm}$-wave pairing and the destructive role of
  apical-oxygen deficiencies in {L}a$_3${N}i$_2${O}$_7$ under pressure}.
\newblock \emph{\bibinfo{journal}{Phys. Rev. Lett.}}
  \textbf{\bibinfo{volume}{131}}, \bibinfo{pages}{236002}
  (\bibinfo{year}{2023}).

\bibitem{zhang2023structural}
\bibinfo{author}{Zhang, Y.}, \bibinfo{author}{Lin, L.-F.},
  \bibinfo{author}{Moreo, A.}, \bibinfo{author}{Maier, T.~A.} \&
  \bibinfo{author}{Dagotto, E.}
\newblock \bibinfo{title}{Structural phase transition, $s_{\pm}$-wave pairing,
  and magnetic stripe order in bilayered superconductor {L}a$_3${N}i$_2${O}$_7$
  under pressure}.
\newblock \emph{\bibinfo{journal}{Nat. Commun.}} \textbf{\bibinfo{volume}{15}},
  \bibinfo{pages}{2470} (\bibinfo{year}{2024}).

\bibitem{zhang2023trends}
\bibinfo{author}{Zhang, Y.}, \bibinfo{author}{Lin, L.-F.},
  \bibinfo{author}{Moreo, A.}, \bibinfo{author}{Maier, T.~A.} \&
  \bibinfo{author}{Dagotto, E.}
\newblock \bibinfo{title}{Trends in electronic structures and $s_{\pm}$-wave
  pairing for the rare-earth series in bilayer nickelate superconductor
  ${R}_3${N}i$_2${O}$_7$}.
\newblock \emph{\bibinfo{journal}{Phys. Rev. B}}
  \textbf{\bibinfo{volume}{108}}, \bibinfo{pages}{165141}
  (\bibinfo{year}{2023}).

\bibitem{gao2025robust}
\bibinfo{author}{Gao, Y.}
\newblock \bibinfo{title}{Robust $s_{\pm}$-wave pairing in a bilayer
  two-orbital model of pressurized {L}a$_3${N}i$_2${O}$_7$ without the $\gamma$
  {F}ermi surface}.
\newblock \emph{\bibinfo{journal}{Physica C}} \textbf{\bibinfo{volume}{640}}, \bibinfo{pages}{1354824}
  (\bibinfo{year}{2026}).

\bibitem{ryee2025optimal}
\bibinfo{author}{Ryee, S.}, \bibinfo{author}{Witt, N.},
  \bibinfo{author}{Sangiovanni, G.} \& \bibinfo{author}{Wehling, T.~O.}
\newblock \bibinfo{title}{Superconductivity governed by {J}anus-faced
  fermiology in strained bilayer nickelates}.
\newblock \emph{\bibinfo{journal}{Phys. Rev. Lett.}}
  \textbf{\bibinfo{volume}{135}}, \bibinfo{pages}{236003}
  (\bibinfo{year}{2025}).

\bibitem{heier2023competing}
\bibinfo{author}{Heier, G.}, \bibinfo{author}{Park, K.} \&
  \bibinfo{author}{Savrasov, S.~Y.}
\newblock \bibinfo{title}{Competing ${d}_{xy}$ and ${s}_{\pm}$ pairing
  symmetries in superconducting {L}a$_3${N}i$_2${O}$_7$:
  $\mathrm{LDA}+\mathrm{FLEX}$ calculations}.
\newblock \emph{\bibinfo{journal}{Phys. Rev. B}}
  \textbf{\bibinfo{volume}{109}}, \bibinfo{pages}{104508}
  (\bibinfo{year}{2024}).

\bibitem{PhysRevB.111.104505}
\bibinfo{author}{Xi, W.}, \bibinfo{author}{Yu, S.-L.} \& \bibinfo{author}{Li,
  J.-X.}
\newblock \bibinfo{title}{Transition from $s_{\pm}$-wave to $d_{x^2-y^2}$-wave
  superconductivity driven by interlayer interaction in the bilayer two-orbital
  model of {L}a$_3${N}i$_2${O}$_7$}.
\newblock \emph{\bibinfo{journal}{Phys. Rev. B}}
  \textbf{\bibinfo{volume}{111}}, \bibinfo{pages}{104505}
  (\bibinfo{year}{2025}).

\bibitem{hua2026possible}
\bibinfo{author}{Hua, Y.}, \bibinfo{author}{He, W.}, \bibinfo{author}{Chen,
  W.-Q.}, \bibinfo{author}{Miao, J.-J.} \& \bibinfo{author}{Yue, C.}
\newblock \bibinfo{title}{Possible enhancement of superconductivity in
  ambient-pressure {L}a$_3${N}i$_2${O}$_7$ thin films}.
\newblock \emph{\bibinfo{journal}{Phys. Rev. B}} \textbf{\bibinfo{volume}{114}}, \bibinfo{pages}{034511} (\bibinfo{year}{2026}).

\bibitem{WangQH2023}
\bibinfo{author}{Yang, Q.-G.}, \bibinfo{author}{Wang, D.} \&
  \bibinfo{author}{Wang, Q.-H.}
\newblock \bibinfo{title}{Possible ${s}_{\pm}$-wave superconductivity in
  {L}a$_3${N}i$_2${O}$_7$}.
\newblock \emph{\bibinfo{journal}{Phys. Rev. B}}
  \textbf{\bibinfo{volume}{108}}, \bibinfo{pages}{L140505}
  (\bibinfo{year}{2023}).

\bibitem{HuJP2023}
\bibinfo{author}{Gu, Y.}, \bibinfo{author}{Le, C.}, \bibinfo{author}{Yang, Z.},
  \bibinfo{author}{Wu, X.} \& \bibinfo{author}{Hu, J.}
\newblock \bibinfo{title}{Effective model and pairing tendency in the bilayer
  {N}i-based superconductor {L}a$_{3}${N}i$_{2}${O}$_{7}$}.
\newblock \emph{\bibinfo{journal}{Phys. Rev. B}}
  \textbf{\bibinfo{volume}{111}}, \bibinfo{pages}{174506}
  (\bibinfo{year}{2025}).

\bibitem{jiang2023high}
\bibinfo{author}{Jiang, K.}, \bibinfo{author}{Wang, Z.} \&
  \bibinfo{author}{Zhang, F.-C.}
\newblock \bibinfo{title}{High-temperature superconductivity in
  {L}a$_3${N}i$_2${O}$_7$}.
\newblock \emph{\bibinfo{journal}{Chin. Phys. Lett.}}
  \textbf{\bibinfo{volume}{41}}, \bibinfo{pages}{017402}
  (\bibinfo{year}{2024}).

\bibitem{fan2023sc}
\bibinfo{author}{Fan, Z.} \emph{et~al.}
\newblock \bibinfo{title}{Superconductivity in nickelate and cuprate
  superconductors with strong bilayer coupling}.
\newblock \emph{\bibinfo{journal}{Phys. Rev. B}}
  \textbf{\bibinfo{volume}{110}}, \bibinfo{pages}{024514}
  (\bibinfo{year}{2024}).

\bibitem{liao2023electron}
\bibinfo{author}{Liao, Z.} \emph{et~al.}
\newblock \bibinfo{title}{Electron correlations and superconductivity in
  {L}a$_{3}${N}i$_{2}${O}$_{7}$ under pressure tuning}.
\newblock \emph{\bibinfo{journal}{Phys. Rev. B}}
  \textbf{\bibinfo{volume}{108}}, \bibinfo{pages}{214522}
  (\bibinfo{year}{2023}).

\bibitem{jiang2023pressure}
\bibinfo{author}{Jiang, R.}, \bibinfo{author}{Hou, J.}, \bibinfo{author}{Fan,
  Z.}, \bibinfo{author}{Lang, Z.-J.} \& \bibinfo{author}{Ku, W.}
\newblock \bibinfo{title}{Pressure driven fractionalization of ionic spins
  results in cupratelike high-${T}_{c}$ superconductivity in
  {L}a$_3${N}i$_2${O}$_7$}.
\newblock \emph{\bibinfo{journal}{Phys. Rev. Lett.}}
  \textbf{\bibinfo{volume}{132}}, \bibinfo{pages}{126503}
  (\bibinfo{year}{2024}).

\bibitem{10.1093/nsr/nwaf353}
\bibinfo{author}{Wang, Z.}, \bibinfo{author}{Zhang, H.-J.},
  \bibinfo{author}{Jiang, K.} \& \bibinfo{author}{Zhang, F.-C.}
\newblock \bibinfo{title}{Self-doped molecular {M}ott insulator for bilayer
  high-temperature superconducting {L}a$_3${N}i$_2${O}$_7$}.
\newblock \emph{\bibinfo{journal}{Natl. Sci. Rev.}}
  \textbf{\bibinfo{volume}{12}}, \bibinfo{pages}{nwaf353}
  (\bibinfo{year}{2025}).

\bibitem{Yi_Feng2023}
\bibinfo{author}{Yang, Y.-F.}, \bibinfo{author}{Zhang, G.-M.} \&
  \bibinfo{author}{Zhang, F.-C.}
\newblock \bibinfo{title}{Interlayer valence bonds and two-component theory for
  high-${T}_{c}$ superconductivity of {L}a$_3${N}i$_2${O}$_7$ under pressure}.
\newblock \emph{\bibinfo{journal}{Phys. Rev. B}}
  \textbf{\bibinfo{volume}{108}}, \bibinfo{pages}{L201108}
  (\bibinfo{year}{2023}).

\bibitem{ZhangGM2023DMRG}
\bibinfo{author}{Shen, Y.}, \bibinfo{author}{Qin, M.} \&
  \bibinfo{author}{Zhang, G.-M.}
\newblock \bibinfo{title}{Effective bi-layer model {H}amiltonian and
  density-matrix renormalization group study for the high-${T}_c$
  superconductivity in {L}a$_3${N}i$_2${O}$_7$ under high pressure}.
\newblock \emph{\bibinfo{journal}{Chin. Phys. Lett.}}
  \textbf{\bibinfo{volume}{40}}, \bibinfo{pages}{127401}
  (\bibinfo{year}{2023}).

\bibitem{kaneko2023pair}
\bibinfo{author}{Kaneko, T.}, \bibinfo{author}{Sakakibara, H.},
  \bibinfo{author}{Ochi, M.} \& \bibinfo{author}{Kuroki, K.}
\newblock \bibinfo{title}{Pair correlations in the two-orbital {H}ubbard
  ladder: Implications for superconductivity in the bilayer nickelate
  {L}a$_3${N}i$_2${O}$_7$}.
\newblock \emph{\bibinfo{journal}{Phys. Rev. B}}
  \textbf{\bibinfo{volume}{109}}, \bibinfo{pages}{045154}
  (\bibinfo{year}{2024}).

\bibitem{WuWei2023charge}
\bibinfo{author}{W{\'u}, W.}, \bibinfo{author}{Luo, Z.}, \bibinfo{author}{Yao,
  D.-X.} \& \bibinfo{author}{Wang, M.}
\newblock \bibinfo{title}{Superexchange and charge transfer in the nickelate
  superconductor {L}a$_3${N}i$_2${O}$_7$ under pressure}.
\newblock \emph{\bibinfo{journal}{Sci. China Phys. Mech. Astron.}}
  \textbf{\bibinfo{volume}{67}}, \bibinfo{pages}{117402}
  (\bibinfo{year}{2024}).

\bibitem{chen2026evolution}
\bibinfo{author}{Chen, Y.}, \bibinfo{author}{Shen, Y.}, \bibinfo{author}{Qian,
  X.}, \bibinfo{author}{Zhang, G.-M.} \& \bibinfo{author}{Qin, M.}
\newblock \bibinfo{title}{Evolution of pairing correlation with $3d_{z^2}$
  electron filling in a bilayer two-orbital model for {L}a$_3${N}i$_2${O}$_7$}.
\newblock \emph{\bibinfo{journal}{Phys. Rev. B}} \textbf{\bibinfo{volume}{114}}, \bibinfo{pages}{245132} (\bibinfo{year}{2026}).

\bibitem{lu2023bilayertJ}
\bibinfo{author}{Lu, C.}, \bibinfo{author}{Pan, Z.}, \bibinfo{author}{Yang, F.}
  \& \bibinfo{author}{Wu, C.}
\newblock \bibinfo{title}{Interlayer-coupling-driven high-temperature
  superconductivity in {L}a$_3${N}i$_2${O}$_7$ under pressure}.
\newblock \emph{\bibinfo{journal}{Phys. Rev. Lett.}}
  \textbf{\bibinfo{volume}{132}}, \bibinfo{pages}{146002}
  (\bibinfo{year}{2024}).

\bibitem{oh2023type2}
\bibinfo{author}{Oh, H.} \& \bibinfo{author}{Zhang, Y.-H.}
\newblock \bibinfo{title}{Type-{II} $t$-${J}$ model and shared superexchange
  coupling from {H}und's rule in superconducting {L}a$_3${N}i$_2${O}$_7$}.
\newblock \emph{\bibinfo{journal}{Phys. Rev. B}}
  \textbf{\bibinfo{volume}{108}}, \bibinfo{pages}{174511}
  (\bibinfo{year}{2023}).

\bibitem{qu2023bilayer}
\bibinfo{author}{Qu, X.-Z.} \emph{et~al.}
\newblock \bibinfo{title}{Bilayer $t$-${J}$-${J}_{\perp}$ model and
  magnetically mediated pairing in the pressurized nickelate
  {L}a$_3${N}i$_2${O}$_7$}.
\newblock \emph{\bibinfo{journal}{Phys. Rev. Lett.}}
  \textbf{\bibinfo{volume}{132}}, \bibinfo{pages}{036502}
  (\bibinfo{year}{2024}).

\bibitem{zhang2023strong}
\bibinfo{author}{Zhang, J.-X.}, \bibinfo{author}{Zhang, H.-K.},
  \bibinfo{author}{You, Y.-Z.} \& \bibinfo{author}{Weng, Z.-Y.}
\newblock \bibinfo{title}{Strong pairing originated from an emergent
  $\mathbb{Z}_2$ {B}erry phase in {L}a$_3${N}i$_2${O}$_7$}.
\newblock \emph{\bibinfo{journal}{Phys. Rev. Lett.}}
  \textbf{\bibinfo{volume}{133}}, \bibinfo{pages}{126501}
  (\bibinfo{year}{2024}).

\bibitem{yang2023strong}
\bibinfo{author}{Yang, H.}, \bibinfo{author}{Oh, H.} \& \bibinfo{author}{Zhang,
  Y.-H.}
\newblock \bibinfo{title}{Strong pairing from a small {F}ermi surface beyond
  weak coupling: Application to {L}a$_3${N}i$_2${O}$_7$}.
\newblock \emph{\bibinfo{journal}{Phys. Rev. B}}
  \textbf{\bibinfo{volume}{110}}, \bibinfo{pages}{104517}
  (\bibinfo{year}{2024}).

\bibitem{wu2024deconfined}
\bibinfo{author}{Wu, X.}, \bibinfo{author}{Yang, H.} \& \bibinfo{author}{Zhang,
  Y.-H.}
\newblock \bibinfo{title}{Deconfined {F}ermi liquid to {F}ermi liquid
  transition and superconducting instability}.
\newblock \emph{\bibinfo{journal}{Phys. Rev. B}}
  \textbf{\bibinfo{volume}{110}}, \bibinfo{pages}{125122}
  (\bibinfo{year}{2024}).

\bibitem{pan2023rno}
\bibinfo{author}{Pan, Z.}, \bibinfo{author}{Lu, C.}, \bibinfo{author}{Yang, F.}
  \& \bibinfo{author}{Wu, C.}
\newblock \bibinfo{title}{Effect of rare-earth element substitution in
  superconducting {R}$_3${N}i$_2${O}$_7$ under pressure}.
\newblock \emph{\bibinfo{journal}{Chin. Phys. Lett.}}
  \textbf{\bibinfo{volume}{41}}, \bibinfo{pages}{087401}
  (\bibinfo{year}{2024}).

\bibitem{Lu2024interplay}
\bibinfo{author}{Lu, C.}, \bibinfo{author}{Pan, Z.}, \bibinfo{author}{Yang, F.}
  \& \bibinfo{author}{Wu, C.}
\newblock \bibinfo{title}{Interplay of two ${E}_{g}$ orbitals in
  superconducting {L}a$_{3}${N}i$_{2}${O}$_{7}$ under pressure}.
\newblock \emph{\bibinfo{journal}{Phys. Rev. B}}
  \textbf{\bibinfo{volume}{110}}, \bibinfo{pages}{094509}
  (\bibinfo{year}{2024}).

\bibitem{PhysRevB.113.174521}
\bibinfo{author}{Chen, Z.}, \bibinfo{author}{Liu, Y.-B.} \&
  \bibinfo{author}{Yang, F.}
\newblock \bibinfo{title}{Variational {M}onte {C}arlo study on the bilayer
  $t$-{$J_{\parallel}$}-{$J_{\perp}$} model for {L}a$_3${N}i$_2${O}$_7$}.
\newblock \emph{\bibinfo{journal}{Phys. Rev. B}}
  \textbf{\bibinfo{volume}{113}}, \bibinfo{pages}{174521}
  (\bibinfo{year}{2026}).

\bibitem{kaneko2025tj}
\bibinfo{author}{Kaneko, T.}, \bibinfo{author}{Kakoi, M.} \&
  \bibinfo{author}{Kuroki, K.}
\newblock \bibinfo{title}{$t$-${J}$ model for strongly correlated two-orbital
  systems: Application to bilayer nickelate superconductors}.
\newblock \emph{\bibinfo{journal}{Phys. Rev. B}}
  \textbf{\bibinfo{volume}{112}}, \bibinfo{pages}{075143}
  (\bibinfo{year}{2025}).

\bibitem{Ji2025StrongCouplingLimit}
\bibinfo{author}{Ji, J.-H.} \emph{et~al.}
\newblock \bibinfo{title}{Strong-coupling study of the pairing
  mechanism in pressurized {L}a$_3${N}i$_2${O}$_7$}.
\newblock \emph{\bibinfo{journal}{Phys. Rev. B}}
  \textbf{\bibinfo{volume}{112}}, \bibinfo{pages}{214515}
  (\bibinfo{year}{2025}).

\bibitem{chen2026unified}
\bibinfo{author}{Chen, Z.}, \bibinfo{author}{Ji, J.-H.}, \bibinfo{author}{Liu,
  Y.-B.}, \bibinfo{author}{Zhang, M.} \& \bibinfo{author}{Yang, F.}
\newblock \bibinfo{title}{Filling and interlayer superexchange control
  superconductivity in {L}a$_3${N}i$_2${O}$_7$}.
\newblock Preprint at \url{https://arxiv.org/abs/2603.14519} (2026).

\bibitem{liu2026triplon}
\bibinfo{author}{Liu, H.} \& \bibinfo{author}{Khaliullin, G.}
\newblock \bibinfo{title}{Triplon-mediated pairing and the superconducting gap
  structure in bilayer nickelates}.
\newblock Preprint at \url{https://arxiv.org/abs/2602.23989} (2026).

\bibitem{shao_electric_field}
\bibinfo{author}{Shao, Z.-Y.}, \bibinfo{author}{Ji, J.-H.},
  \bibinfo{author}{Wu, C.}, \bibinfo{author}{Yao, D.-X.} \&
  \bibinfo{author}{Yang, F.}
\newblock \bibinfo{title}{Possible liquid-nitrogen-temperature
  superconductivity driven by perpendicular electric field in the
  single-bilayer film of {L}a$_3${N}i$_2${O}$_7$ at ambient pressure}.
\newblock \emph{\bibinfo{journal}{Nat. Commun.}} \textbf{\bibinfo{volume}{17}},
  \bibinfo{pages}{1120} (\bibinfo{year}{2026}).

\bibitem{fan2025minimal}
\bibinfo{author}{Fan, Z.-D.} \& \bibinfo{author}{Vishwanath, A.}
\newblock \bibinfo{title}{Minimal two band model and experimental proposals to
  distinguish pairing mechanisms of the high-{T}$_c$ superconductor
  {L}a$_3${N}i$_2${O}$_7$}.
\newblock Preprint at \url{https://arxiv.org/abs/2512.05956} (2025).

\bibitem{qiu2025pairing}
\bibinfo{author}{Qiu, W.}, \bibinfo{author}{Luo, Z.}, \bibinfo{author}{Hu, X.}
  \& \bibinfo{author}{Yao, D.-X.}
\newblock \bibinfo{title}{Pairing symmetry and superconductivity in
  {L}a$_3${N}i$_2${O}$_7$ thin films}.
\newblock \emph{\bibinfo{journal}{Chin. Phys. Lett.}} \textbf{\bibinfo{volume}{43}}, \bibinfo{pages}{080710} (\bibinfo{year}{2026}).

\bibitem{PhysRevB.18.4945}
\bibinfo{author}{Castellani, C.}, \bibinfo{author}{Natoli, C.~R.} \&
  \bibinfo{author}{Ranninger, J.}
\newblock \bibinfo{title}{Magnetic structure of
  ${\mathrm{V}}_{2}$${\mathrm{O}}_{3}$ in the insulating phase}.
\newblock \emph{\bibinfo{journal}{Phys. Rev. B}} \textbf{\bibinfo{volume}{18}},
  \bibinfo{pages}{4945--4966} (\bibinfo{year}{1978}).

\bibitem{takimoto2004strong}
\bibinfo{author}{Takimoto, T.}, \bibinfo{author}{Hotta, T.} \&
  \bibinfo{author}{Ueda, K.}
\newblock \bibinfo{title}{Strong-coupling theory of superconductivity in a
  degenerate {H}ubbard model}.
\newblock \emph{\bibinfo{journal}{Phys. Rev. B}} \textbf{\bibinfo{volume}{69}},
  \bibinfo{pages}{104504} (\bibinfo{year}{2004}).

\bibitem{yada2005origin}
\bibinfo{author}{Yada, K.} \& \bibinfo{author}{Kontani, H.}
\newblock \bibinfo{title}{Origin of weak pseudogap behaviors in
  {N}a$_{0.35}${C}o{O}$_2$: Absence of small hole pockets}.
\newblock \emph{\bibinfo{journal}{J. Phys. Soc. Jpn.}}
  \textbf{\bibinfo{volume}{74}}, \bibinfo{pages}{2161--2164}
  (\bibinfo{year}{2005}).

\bibitem{kubo2007pairing}
\bibinfo{author}{Kubo, K.}
\newblock \bibinfo{title}{Pairing symmetry in a two-orbital {H}ubbard model on a
  square lattice}.
\newblock \emph{\bibinfo{journal}{Phys. Rev. B}} \textbf{\bibinfo{volume}{75}},
  \bibinfo{pages}{224509} (\bibinfo{year}{2007}).

\bibitem{graser2009near}
\bibinfo{author}{Graser, S.}, \bibinfo{author}{Maier, T.~A.},
  \bibinfo{author}{Hirschfeld, P.~J.} \& \bibinfo{author}{Scalapino, D.~J.}
\newblock \bibinfo{title}{Near-degeneracy of several pairing channels in
  multiorbital models for the {F}e pnictides}.
\newblock \emph{\bibinfo{journal}{New J. Phys.}} \textbf{\bibinfo{volume}{11}},
  \bibinfo{pages}{025016} (\bibinfo{year}{2009}).

\bibitem{liu2013d+}
\bibinfo{author}{Liu, F.}, \bibinfo{author}{Liu, C.-C.}, \bibinfo{author}{Wu,
  K.}, \bibinfo{author}{Yang, F.} \& \bibinfo{author}{Yao, Y.}
\newblock \bibinfo{title}{$d+ id'$ chiral superconductivity in bilayer
  silicene}.
\newblock \emph{\bibinfo{journal}{Phys. Rev. Lett.}}
  \textbf{\bibinfo{volume}{111}}, \bibinfo{pages}{066804}
  (\bibinfo{year}{2013}).

\bibitem{zhang2022lifshitz}
\bibinfo{author}{Zhang, M.}, \bibinfo{author}{Hao, J.-J.}, \bibinfo{author}{Wu,
  X.} \& \bibinfo{author}{Yang, F.}
\newblock \bibinfo{title}{Lifshitz transition enhanced triplet $p_z$-wave
  superconductivity in hydrogen-doped {K}{C}r$_3${A}s$_3$}.
\newblock \emph{\bibinfo{journal}{Phys. Rev. B}}
  \textbf{\bibinfo{volume}{105}}, \bibinfo{pages}{134509}
  (\bibinfo{year}{2022}).

\bibitem{kuroki101unconventional}
\bibinfo{author}{Kuroki, K.} \emph{et~al.}
\newblock \bibinfo{title}{Unconventional pairing originating from the
  disconnected {F}ermi surfaces of superconducting
  {L}a{F}e{A}s{O}$_{1-x}${F}$_x$}.
\newblock \emph{\bibinfo{journal}{Phys. Rev. Lett.}}
  \textbf{\bibinfo{volume}{101}}, \bibinfo{pages}{087004}
  (\bibinfo{year}{2008}).

\bibitem{cao2026tunable}
\bibinfo{author}{Cao, Y.-H.}, \bibinfo{author}{Jiang, K.-Y.},
  \bibinfo{author}{Lu, H.-Y.}, \bibinfo{author}{Wang, D.} \&
  \bibinfo{author}{Wang, Q.-H.}
\newblock \bibinfo{title}{Tunable superconductivity and spin density waves in
  {La}$_3${Ni}$_2${O}$_7$/{LaAlO}$_3$ thin films}.
\newblock \emph{\bibinfo{journal}{Phys. Rev. B}}
  \textbf{\bibinfo{volume}{114}}, \bibinfo{pages}{154511}
  (\bibinfo{year}{2026}).

\bibitem{wu2026interorbital}
\bibinfo{author}{Wu, X.}, \bibinfo{author}{Xiang, T.} \& \bibinfo{author}{Hu, J.}
\newblock \bibinfo{title}{Pairing mechanism in bilayer nickelate {La}$_3${Ni}$_2${O}$_7$ superconductors}.
\newblock Preprint at \url{https://arxiv.org/abs/2604.17181} (2026).

\bibitem{li2025substrate}
Li, G. \emph{et al.}
\newblock Theoretical study on the electronic properties and multiorbital models of La$_3$Ni$_2$O$_7$ thin films on SrLaAlO$_4$ (001).
\newblock \emph{AAPPS Bull.} \textbf{36}, 13 (2026).

\bibitem{scholten2026nodal}
Scholten, M., B{\"o}tzel, S., Lechermann, F., Fernandes, R. M. \& Eremin, I. M.
\newblock Nodal orbital-anti-phase superconducting state in bilayer nickelates.
\newblock Preprint at \url{https://arxiv.org/abs/2609.29263} (2026).

\end{thebibliography}

\begin{thebibliography}{7}
\expandafter\ifx\csname url\endcsname\relax
  \def\url#1{\texttt{#1}}\fi
\expandafter\ifx\csname urlprefix\endcsname\relax\def\urlprefix{URL }\fi
\providecommand{\bibinfo}[2]{#2}
\providecommand{\eprint}[2][]{\url{#2}}

\bibitem{sm-yue2025correlated}
\bibinfo{author}{Yue, C.} \emph{et~al.}
\newblock \bibinfo{title}{Correlated electronic structures and unconventional
  superconductivity in bilayer nickelate heterostructures}.
\newblock \emph{\bibinfo{journal}{Natl. Sci. Rev.}}
  \textbf{\bibinfo{volume}{12}}, \bibinfo{pages}{nwaf253}
  (\bibinfo{year}{2025}).

\bibitem{sm-Daoxin_Yao2025}
\bibinfo{author}{Hu, X.}, \bibinfo{author}{Qiu, W.}, \bibinfo{author}{Chen,
  C.-Q.}, \bibinfo{author}{Luo, Z.} \& \bibinfo{author}{Yao, D.-X.}
\newblock \bibinfo{title}{Electronic structures and multi-orbital models of
  {L}a$_3${N}i$_2${O}$_7$ thin films at ambient pressure}.
\newblock \emph{\bibinfo{journal}{Commun. Phys.}}
  \textbf{\bibinfo{volume}{8}}, \bibinfo{pages}{506} (\bibinfo{year}{2025}).

\bibitem{sm-cao2026tunable}
\bibinfo{author}{Cao, Y.-H.}, \bibinfo{author}{Jiang, K.-Y.},
  \bibinfo{author}{Lu, H.-Y.}, \bibinfo{author}{Wang, D.} \&
  \bibinfo{author}{Wang, Q.-H.}
\newblock \bibinfo{title}{Tunable superconductivity and spin density waves in
  {La}$_3${Ni}$_2${O}$_7$/{LaAlO}$_3$ thin films}.
\newblock \emph{\bibinfo{journal}{Phys. Rev. B}}
  \textbf{\bibinfo{volume}{114}}, \bibinfo{pages}{154511}
  (\bibinfo{year}{2026}).

\bibitem{sm-cao2025strain}
\bibinfo{author}{Cao, Y.-H.}, \bibinfo{author}{Jiang, K.-Y.},
  \bibinfo{author}{Lu, H.-Y.}, \bibinfo{author}{Wang, D.} \&
  \bibinfo{author}{Wang, Q.-H.}
\newblock \bibinfo{title}{Strain-engineered electronic structure and
  superconductivity in {L}a$_3${N}i$_2${O}$_7$ thin films}.
\newblock \emph{\bibinfo{journal}{Sci. China Phys. Mech. Astron.}}
  \textbf{\bibinfo{volume}{69}}, \bibinfo{pages}{247412}
  (\bibinfo{year}{2026}).

\bibitem{sm-ryee2025optimal}
\bibinfo{author}{Ryee, S.}, \bibinfo{author}{Witt, N.},
  \bibinfo{author}{Sangiovanni, G.} \& \bibinfo{author}{Wehling, T.~O.}
\newblock \bibinfo{title}{Superconductivity governed by {J}anus-faced
  fermiology in strained bilayer nickelates}.
\newblock \emph{\bibinfo{journal}{Phys. Rev. Lett.}}
  \textbf{\bibinfo{volume}{135}}, \bibinfo{pages}{236003}
  (\bibinfo{year}{2025}).

\bibitem{sm-li2025substrate}
Li, G. \emph{et al.}
\newblock Theoretical study on the electronic properties and multiorbital models of La$_3$Ni$_2$O$_7$ thin films on SrLaAlO$_4$ (001).
\newblock \emph{AAPPS Bull.} \textbf{36}, 13 (2026).

\bibitem{sm-shen2025anomalous}
\bibinfo{author}{Shen, J.} \emph{et~al.}
\newblock \bibinfo{title}{Nodeless superconducting gap and electron-boson
  coupling in ({L}a,{P}r,{S}m)$_3${N}i$_2${O}$_7$ films}.
\newblock \emph{\bibinfo{journal}{Science}} \textbf{\bibinfo{volume}{392}},
  \bibinfo{pages}{1396--1400} (\bibinfo{year}{2026}).

\end{thebibliography}
\end{document}